\documentclass[prb,twocolumn,showpacs]{revtex4-1}
\usepackage[utf8x]{inputenc}
\usepackage{amsbsy}
\usepackage{graphicx}
\usepackage{color}
\usepackage{dcolumn}
\usepackage{amsmath}
\usepackage{amssymb}
\usepackage{amsfonts}
\usepackage{subfigure}

\usepackage[colorlinks=true, citecolor=black, linkcolor=black, urlcolor=black]{hyperref}

\newcommand{\abinitio}{\emph{ab initio}}
\newcommand{\etal}{\emph{et al.}}

\newcommand{\bra}[1]{\langle #1|}
\newcommand{\ket}[1]{|#1 \rangle }
\newcommand{\beq}{\begin{equation}}
\newcommand{\eeq}{\end{equation}}
\newcommand{\beqn}{\begin{eqnarray}}
\newcommand{\eeqn}{\end{eqnarray}}

\begin{document}
\title{Anharmonic free energies and phonon dispersions from the stochastic 
       self-consistent harmonic approximation: application to
       platinum and palladium hydrides}

\author{Ion Errea$^{1, 2}$}
\author{Matteo Calandra$^{1}$}
\author{Francesco Mauri$^{1}$}

\affiliation{$^1$Universit\'e Pierre et Marie Curie-Paris 6, CNRS, IMPMC-UMR7590,
                 case 115, 4 Place Jussieu, 75252 Paris Cedex 05, France}
\affiliation{$^2$IKERBASQUE, Basque Foundation for Science, 48011, Bilbao, Spain}

\begin{abstract}
Harmonic calculations based on density-functional theory are generally
the method of choice for the description of phonon spectra 
of metals and insulators. The inclusion of anharmonic effects 
is, however, delicate as it relies on perturbation theory requiring 
a considerable amount of computer time, fast increasing with the 
cell size. Furthermore, perturbation theory
breaks down when the harmonic solution is dynamically unstable
or the anharmonic correction of the phonon energies is larger than the
harmonic frequencies themselves.
We present here a stochastic implementation of the self-consistent
harmonic approximation valid to treat anharmonicity at any temperature 
in the non-perturbative regime.
The method is based on the minimization of the free energy with respect
to a trial density matrix described by an arbitrary harmonic Hamiltonian.
The minimization is performed with respect to all the free parameters in the
trial harmonic Hamiltonian, namely, equilibrium positions, phonon frequencies
and polarization vectors. The gradient of the free energy is 
calculated following a stochastic procedure. The method can be used to
calculate thermodynamic properties, dynamical properties and even anharmonic 
corrections to the Eliashberg function of the 
electron-phonon coupling. The scaling with the
system size is greatly improved with respect to perturbation theory.
The validity of the method is demonstrated
in the strongly anharmonic palladium and platinum hydrides.
In both cases we predict a strong
anharmonic correction to the harmonic phonon spectra, far beyond the perturbative limit. 
In palladium hydrides we calculate thermodynamic properties
beyond the quasiharmonic approximation, while in PtH we demonstrate that
the high superconducting critical temperatures at 100 GPa predicted in
previous calculations based on the harmonic approximation are strongly
suppressed when anharmonic effects are included. 
\end{abstract}

%\pacs{74.25.Kc,74.90.+n,67.63.-r,67.80.F-,63.20.Ry,31.15.A-}

\maketitle

%%%%%%%%%%%%%%%% 
% Introduction %
%%%%%%%%%%%%%%%%

\section{Introduction}
\label{introduction}

Describing accurately vibrations of atoms is of paramount importance
in the physical properties of solids, liquids and molecules. 
The contribution of atomic vibrations to the free energy of the system 
affects equilibrium and thermodynamic properties, 
while the frequencies and the deformation potentials 
determine transport and superconducting properties.
Moreover, the spectra obtained from spectroscopic
techniques such as infrared, Raman, and inelastic X-ray or  
neutron scattering cannot be understood without accounting for atomic vibrations.
The quantum mechanical description of atomic vibrations in terms of 
phonons or vibrons 
has provided a successful framework to describe all these
properties~\cite{born-huang}. 

Nowadays, vibron energies and phonon dispersions in the 
harmonic approximation are routinely calculated from first-principles 
making use of linear response theory~\cite{RevModPhys.73.515}
or the small displacement method~\cite{0295-5075-32-9-005}, and 
thermodynamic properties can be accounted within the standard
quasiharmonic approximation~\cite{Fultz2010247}. The harmonic 
approximation relies on the following assumptions: (i)
the displacement of the atoms from their equilibrium positions
is small compared to the interatomic distance and, as a consequence, 
(ii) the ionic potential can be approximated
with the truncation at second order of the 
Born-Oppenheimer (BO) energy surface.
The harmonic approximation predicts
that phonons or vibrons are well-defined quasiparticles with an infinite lifetime.
Thus, finite values of the thermal conductivity in solids 
cannot be accounted for.
Moreover, harmonic vibrational energies are temperature
independent and, therefore,
thermodynamic properties at high temperature might not
be properly accounted within the harmonic approximation. 
Despite being computationally challenging for \abinitio{} 
approaches,
phonons finite lifetime and the temperature dependence
of their frequencies can be explained treating third and 
fourth order terms in the expansion of the BO energy surface
within perturbation 
theory~\cite{PhysRev.128.2589,PhysRevB.68.220509,
Lazzeri,PhysRevB.82.104504,PhysRevB.87.214303} at a very high
computational cost, fast increasing with the system size.
  
The validity of perturbation theory is, however, limited to situations in which
the harmonic potential is considerably larger than higher order terms. 
Then, the perturbative correction of vibrational 
frequencies is small with respect to the
harmonic result. Nevertheless, whenever the displacements of the
atoms largely exceed the range in which the harmonic potential is valid,
the harmonic approximation and any perturbative approach based on it
break down~\cite{hooton422,1742-6596-377-1-012060}. 
This situation occurs whenever a system is close to a dynamical instability,  
light atoms are present, or temperature
is high and the solid is not far from melting. 
In these cases, describing the temperature dependence of the phonon
spectra is crucial and incorporating anharmonic corrections to the
free energy is mandatory to describe properly thermodynamic
properties. This non-perturbative regime has already been identified
in superconductors~\cite{PhysRevLett.106.165501,PhysRevLett.96.047003,
PhysRevLett.77.1151,PhysRevLett.99.155505,PhysRevB.74.094519},
transition-metal dichalcogenides with charge-density
waves~\cite{PhysRevB.86.155125,
PhysRevB.80.241108,PhysRevLett.106.196406,PhysRevB.77.165135,PhysRevB.73.205102}, 
thermoelectric materials~\cite{delaire614}, 
ferroelectrics~\cite{PhysRevB.52.6301,PhysRevLett.74.4067},
hydrides~\cite{PhysRevLett.111.177002}, materials under extreme 
temperature or pressure conditions~\cite{vocadlo536,Luo01062010,
RevModPhys.84.945},
or in the isotopic fractionation of water~\cite{Markland22052012} 
to mention but a few examples.    
The development of
a non-perturbative treatment of phonon-phonon scattering is thus a
major challenge for many fields of physics and chemistry.

\emph{Ab initio} molecular dynamics (AIMD)~\cite{PhysRevLett.55.2471}
calculations are the most common way of treating anharmonicity
at any order. However, as they are based on Newtonian mechanics,
quantum effects important at low temperature are not properly
characterized by AIMD, and, consequently, the application of
AIMD is limited to temperatures
above Debye temperature. 
The quantum behavior can be incorporated to AIMD making use of
quantum thermal baths~\cite{PhysRevLett.103.190601,PhysRevLett.103.030603}.
However, this approach is exclusively valid for harmonic
potentials~\cite{PhysRevLett.107.198901,PhysRevLett.107.198902}.
The problem is overcome by path-integral
molecular dynamics (PIMD)~\cite{RevModPhys.67.279}, but 
the great computational cost of the method makes it 
challenging for density-functional theory (DFT)
state-of-the-art calculations. 

Aiming to overcome these difficulties, several methods
have been developed recently to deal with anharmonic 
effects beyond perturbation 
theory~\cite{PhysRevLett.106.165501,PhysRevLett.100.095901,Souvatzis2009888,
PhysRevB.86.054119,PhysRevB.84.180301,
PhysRevB.87.104111,PhysRevB.87.144302} mainly inspired by the self-consistent
harmonic approximation (SCHA) devised by Hooton~\cite{hooton422}.
The main idea of the SCHA is that the system can be approximated
by the harmonic potential that minimizes the free energy of
the system, which does not necessarily coincide with 
the potential obtained from the second derivatives 
of the BO energy surface. 
The self-consistent \abinitio{} lattice dynamics (SCAILD) 
method~\cite{PhysRevLett.100.095901,Souvatzis2009888}
is an iterative way of converging the phonon frequencies 
at different temperatures accounting for anharmonic effects, 
but it does not optimize the 
eigenvectors of the harmonic potential nor the internal
parameters in the crystal or molecular structure, and does not
include anharmonic corrections in the free energy.
The method presented by Antolin \etal{} does not include
such corrections either~\cite{PhysRevB.86.054119}.
The temperature dependent effective potential
(TDEP) technique can optimize the potential
with respect to both polarization vectors and internal parameters,
and, in principle,
can include anharmonic corrections to the free energy
through thermodynamic 
integration~\cite{PhysRevB.84.180301,PhysRevB.87.104111}.
Nevertheless, as it is based on AIMD calculations, it might break down
below Debye temperature and the thermodynamic 
integration technique might be inefficient in their 
scheme~\cite{PhysRevB.87.104111}.
Finally, Monserrat \etal{} have recently presented a method 
in which the BO energy surface is Taylor expanded as a function
of the harmonic normal coordinates~\cite{PhysRevB.87.144302}. 
The obtained Hamiltonian is solved variationally. 
In this scheme, the internal parameters of the crystal
structure are not optimized, and the required 
mapping of the BO energy surface might demand
a large computational effort. 

In this paper we present a new implementation of the SCHA that
is fully variational in the free energy. The
free energy is explicitly minimized using a conjugate-gradient
(CG) algorithm with respect to all the
independent coefficients in a trial harmonic potential. 
Therefore, the method allows to 
access directly the anharmonic free energy of the system and
optimizes the free energy with respect to phonon or vibron frequencies,
polarization vectors and free parameters in the crystal or molecular
structure. The temperature dependence is naturally incorporated into the formalism
and temperature-dependent phonon dispersions or vibron 
frequencies can be readily calculated.
The method is based on a stochastic evaluation of the free energy
and its gradient. Thus, the cumbersome evaluation of anharmonic 
forces~\cite{PhysRevLett.106.165501,errea:112604} 
or mapping the BO energy surface
is avoided. The method is named as the stochastic self-consistent
harmonic approximation (SSCHA). 
It is perfectly valid to study both lattice or molecular vibrations, 
but, in order to simplify the text, we will use the language of
crystals throughout. 
The SSCHA requires the calculation of total energies and forces
on supercells with suitably chosen ionic configurations, which
can be computed at any degree of theory. The SSCHA algorithm 
is devised to minimize the number of total energy and force
calculations. 

We apply the method to the strongly anharmonic platinum and palladium 
hydrides. In both cases the anharmonic
correction to the phonon frequencies is larger than the
harmonic frequencies themselves, invalidating any perturbative approach.  
We first study the
role of anharmonicity in PtH at high pressure fully from first-principles, 
demonstrating that
the high superconducting critical temperatures predicted in previous works
~\cite{PhysRevLett.107.117002,PhysRevB.83.214106,PhysRevB.84.054543} 
are strongly suppressed by
anharmonic effects. This result questions the interpretation 
suggested by several authors~\cite{PhysRevB.83.214106,
PhysRevB.84.054543,PhysRevLett.107.117002} of the experiment in 
silane by Eremets \etal{}~\cite{Eremets14032008}, 
where superconductivity was measured for the first time in a 
high-pressure hydride, stating that the measured 
superconductivity corresponded not to silane but to PtH. 
In palladium hydrides we show how within the SSCHA we can 
calculate thermodynamic properties in agreement with experiments 
in cases where the quasiharmonic approximation 
breaks down.

The paper is structured as follows. In Sec. \ref{scha} we present the
theoretical foundation of the SCHA and in Sec. \ref{sscha}
the way we implement it in a stochastic manner. In Secs.
\ref{pth} and \ref{pdh}
we apply the SSCHA to the strongly anharmonic platinum and palladium
hydrides, 
where no perturbative approach is feasible. 
Finally, summary and conclusions are given in Sec. \ref{conclusions}.   

%%%%%%%%%%%%
% The SCHA %
%%%%%%%%%%%%

\section{The self-consistent harmonic approximation}
\label{scha}

The SSCHA method applies to molecules and solids.
In terms of notation clarity we treat the 
physical system as an isolated molecule throughout the manuscript.
This means that in the case of periodic crystals we take a periodic 
supercell and treat the system at the $\Gamma$ point. 
As it will be explained in Sec. \ref{parameters}, in the
latter case we take advantage of translational symmetries.

Within the BO or adiabatic approximation, which assumes that the
electrons adapt instantaneously to the ionic positions, the
dynamics of the ions in the supercell are determined by $V$,
the potential defined by the BO energy surface.
Normally, this potential is Taylor expanded as a function of the
ionic displacements as
\beq
V = V_0 + \sum_{n=2}^{\infty} V_n,
\label{potential-taylor}
\eeq
where
\beq
V_n = \frac{1}{n!} \sum_{s_1 \dots s_n} 
        \sum_{\alpha_1 \dots \alpha_n} \phi_{s_1 \dots s_n}^{\alpha_1 \dots \alpha_n}
        u^{s_1 \alpha_1} \dots  u^{s_n \alpha_n} 
\label{n-order-potential}
\eeq
and 
\beq
u^{s \alpha} = R^{s \alpha} - R_{{\rm eq}}^{s \alpha}
\label{u-operator}
\eeq
is the out-of-equilibrium displacement of atom $s$ in the supercell 
along Cartesian coordinate $\alpha$, with
$R^{s \alpha}$ the corresponding atomic position  
and $R_{{\rm eq}}^{s \alpha}$ the atomic equilibrium position. In Eq. 
\eqref{n-order-potential} $ \phi_{s_1 \dots s_n}^{\alpha_1 \dots \alpha_n}$
represents the $n$-th order derivative of the BO energy surface 
with respect to the atomic displacements calculated at
equilibrium: 
\beq
\phi_{s_1 \dots s_n}^{\alpha_1 \dots \alpha_n} = \left[ \frac{ \partial^{(n)} 
  V }{ \partial u^{s_1 \alpha_1} \dots \partial u^{s_n \alpha_n}}
  \right]_0.  
\label{derivative-potential}
\eeq
Note that in Eq. \eqref{potential-taylor} the first order term in the expansion
vanishes as forces are zero at the equilibrium position.

Once the potential is defined, the dynamics of the ionic degrees of freedom
are determined by the 
\beq
H = T + V
\label{hamiltonian-total}
\eeq 
Hamiltonian, where
\beq
T = \sum_{s=1}^{N} \sum_{\alpha=1}^3 \frac{ 
            ( P^{s \alpha} )^2 }{2 M_s}
\label{kinetic}
\eeq
is the kinetic-energy operator of the ions, with $N$ the total number
of atoms in the supercell, $P^{s \alpha}$ the momentum operator of the $s$-th
atom along $\alpha$, and $M_s$ the mass of the $s$-th atom. 

%
%---------------------------
%

\subsection{Formal definition of the self-consistent harmonic approximation}
\label{scha-def}

The free energy of the ionic Hamiltonian is given by 
the sum of the total energy and the entropic contribution:
\beq
F_H = \mathrm{tr} [ \rho_H H] + \frac{1}{\beta} 
      \mathrm{tr} [ \rho_H \ln \rho_H] = 
      - \frac{1}{\beta} \ln Z_H,
\label{true-free-energy}
\eeq
where the partition function is $Z_H = \mathrm{tr} [ e^{- \beta H}]$,
the density matrix $\rho_H = e^{- \beta H} / Z_H$, and
$\beta = 1 / (k_BT)$. Calculating $F_H$ represents
a complicated task due to the many-body character of $V$.
Instead, a quantum variational principle in the free energy can be defined
substituting the density matrix by any density matrix 
$\rho_{\mathcal{H}}$ defined by a trial 
$\mathcal{H} = T + \mathcal{V}$ Hamiltonian. 
Then, if
\beq
\mathcal{F}_H[\mathcal{H}] = \mathrm{tr} [ \rho_{\mathcal{H}} H] + 
      \frac{1}{\beta} 
      \mathrm{tr} [ \rho_{\mathcal{H}} \ln \rho_{\mathcal{H}}]
\label{var-free-energy}
\eeq
we have the so-called Gibbs-Bogoliubov inequality~\cite{0022-3689-1-5-305},
namely 
\beq
F_H \le \mathcal{F}_H[\mathcal{H}].
\label{gibbs-bogoliubov}
\eeq
Adding and subtracting $\mathrm{tr} [ \rho_{\mathcal{H}} \mathcal{H}]$
in Eq. \eqref{var-free-energy}, it is straightforward to demonstrate that
\beq
\mathcal{F}_H[\mathcal{H}] = F_{\mathcal{H}} + \mathrm{tr} [ \rho_{\mathcal{H}}
      ( V - \mathcal{V}) ]. 
\label{var-free-energy2}
\eeq
Obviously, the equality holds in Eq. \eqref{gibbs-bogoliubov}
when $H = \mathcal{H}$. Thus, if
$\mathcal{F}_H[\mathcal{H}]$ is minimized with respect to the
trial $\mathcal{H}$ Hamiltonian, a quantum variational 
principle is established valid at any temperature for the
ionic problem.

The SCHA, which was originally
proposed by Hooton~\cite{hooton422} and it was further 
developed by Choquard~\cite{choquard} and Werthamer~\cite{PhysRevB.1.572}, 
is obtained by restricting the trial potential $ \mathcal{V}$ to
a harmonic one. The SCHA is analogous to the Hartree-Fock
approximation for electrons in the sense that it assumes 
a trial density matrix formed by 
single-particle wave functions.
As we shall see in Sec. \ref{harmonic},
one advantage of taking a harmonic potential is that
$F_{\mathcal{H}}$ and the probability density
to find the system in a general $\mathbf{R}$ 
ionic configuration, 
$\rho_{\mathcal{H}}(\mathbf{R}) = \bra{\mathbf{R}} 
\rho_{\mathcal{H}} \ket{\mathbf{R}}$,
can be expressed in a closed form in terms of phonon
frequencies, polarization vectors and equilibrium positions.

The variational principle proposed by the SCHA 
allows us to treat systems beyond
perturbation theory since, 
even if $V - V_2$ 
is large compared to $V_2$ itself
invalidating any perturbative 
approach~\cite{PhysRev.128.2589,PhysRevB.68.220509,
Lazzeri,PhysRevB.82.104504,PhysRevB.87.214303},
the variational principle is still valid. 
It should be remarked that in systems with huge anharmonicity 
phonons remain well-defined quasiparticles that are actually
measured~\cite{PhysRevLett.106.165501,PhysRevLett.96.047003,
PhysRevLett.77.1151,PhysRevLett.99.155505,PhysRevB.74.094519,PhysRevB.86.155125,
PhysRevB.80.241108,PhysRevLett.106.196406,PhysRevB.77.165135,PhysRevB.73.205102, 
delaire614,PhysRevB.52.6301,PhysRevLett.74.4067,
PhysRevLett.111.177002,vocadlo536,Luo01062010,
RevModPhys.84.945,Markland22052012}.
Similarly, electrons in solids are existing quasiparticles
despite being strongly affected by the 
electron-electron Coulomb interaction. 
Therefore, once the harmonic potential $\mathcal{V}$ that 
minimizes $\mathcal{F}_H[\mathcal{H}]$
has been found, the eigenvalues of $\mathcal{V}$ 
can be ascribed to the phonon spectra
renormalized by anharmonic effects. 

%
%---------------------------
%

\subsection{The harmonic Hamiltonian}
\label{harmonic}

The trial harmonic Hamiltonian of the SCHA 
is written in its general 
\beq
\mathcal{H} = \sum_{s=1}^{N} \sum_{\alpha=1}^3 \frac{ 
            ( P^{s \alpha} )^2 }{2 M_s}
            + \frac{1}{2} \sum_{st}^N \sum_{\alpha \beta}^3 
               u^{s \alpha} \Phi_{st}^{\alpha \beta}
              u^{t \beta}
\label{harmonic-potential}
\eeq
form. The trial force-constants matrix $\Phi_{st}^{\alpha \beta}$
is different from the force-constants matrix
associated to $V_2$, $\phi_{st}^{\alpha \beta}$. 
Diagonalizing the dynamical matrix $\Phi_{st}^{\alpha \beta}/\sqrt{M_sM_t}$ as
\beq
\sum_{t=1}^{N} \sum_{\beta=1}^3 \frac{\Phi_{st}^{\alpha \beta}}{\sqrt{M_sM_t}}
             \epsilon_{\mu \mathcal{H}}^{t \beta} = 
             \omega^2_{\mu \mathcal{H}} \epsilon_{\mu \mathcal{H}}^{s \alpha},
\label{eigenvalue-eq}
\eeq
the polarization vectors $\epsilon_{\mu \mathcal{H}}^{s \alpha}$
and the $\omega_{\mu \mathcal{H}}$ phonon frequncies are obtained. 
These allow us to define the $q_{\mu}$ and $p_{\mu}$ 
normal coordinates that transform as
\beqn
u^{s \alpha} & = & \sum_{\mu=1}^{3N} \frac{1}{\sqrt{M_s}} 
                         \epsilon_{\mu \mathcal{H}}^{s \alpha} q_{\mu}
                         \label{qmu} \\
P^{s \alpha} & = & \sum_{\mu=1}^{3N} \sqrt{M_s} 
                         \epsilon_{\mu \mathcal{H}}^{s \alpha} p_{\mu}.
                         \label{pmu} 
\eeqn
Applying the change of variables in Eqs. \eqref{qmu} and \eqref{pmu}
to Eq. \eqref{harmonic-potential}, 
$\mathcal{H}$ can be written as a sum of $3N$
independent oscillators:
\beq
\mathcal{H} = \sum_{\mu=1}^{3N} \left( \frac{p_{\mu}^2}{2} +
                                     \frac{\omega^2_{\mu \mathcal{H}} 
                                     q_{\mu}^2}{2}
                              \right).
\label{harmonic-potential-normal}
\eeq
In Eqs. \eqref{eigenvalue-eq}-\eqref{harmonic-potential-normal}
$\mu$ is a mode index and the subindex $\mathcal{H}$ 
in the phonon frequencies and polarization vectors
denotes that they are associated to the 
harmonic Hamiltonian $\mathcal{H}$.

Once $\mathcal{H}$ has been written as a sum of $3N$ independent harmonic 
oscillators, it is easy to observe that the
probability to find the system in a general ionic configuration 
$\mathbf{R}$ is (see Appendix \ref{app-gradient})
\beq
\rho_{\mathcal{H}}(\mathbf{R}) = A_{\mathcal{H}} 
     \exp \left[ -  \sum\limits_{st \alpha \beta \mu} \frac{ \sqrt{M_sM_t} }{ 
     2 a_{\mu \mathcal{H}}^2}
 \epsilon_{\mu \mathcal{H}}^{s\alpha}
                \epsilon_{\mu \mathcal{H}}^{t\beta} u^{s\alpha} u^{t\beta} 
    \right],
\label{probability}
\eeq
where $A_{\mathcal{H}}$ is the normalization constant
and 
\beq
a_{\mu \mathcal{H}}=\sqrt{\hbar \coth (\beta \hbar \omega_{\mu
\mathcal{H}} / 2) / (2 \omega_{\mu \mathcal{H}})}
\label{normal-length}
\eeq
is called the normal length of mode $\mu$, even if it has units 
of length times square root of mass.
Then, the quantum
statistical average of any observable $O$
that is exclusively a function of the atomic positions
can be computed as
\beq
\mathrm{tr} [\rho_{\mathcal{H}} O] 
         = \int \mathrm{d}\mathbf{R} O(\mathbf{R}) \rho_{\mathcal{H}}(\mathbf{R}).
\label{quantum-average}
\eeq
Moreover, for a harmonic Hamiltonian its free energy can be calculated 
analytically from the well-known
\beq
F_{\mathcal{H}} = \sum_{\mu=1}^{3N} \left[ \frac{1}{2} \hbar \omega_{\mu \mathcal{H}} 
    - \frac{1}{\beta} \mathrm{ln} [1 + n_B(\omega_{\mu \mathcal{H}})] \right] 
\label{f0}
\eeq
equation, where $n_B(\omega) = 1 / \left( e^{\beta \hbar \omega} - 1\right)$
is the bosonic occupation factor. 
The fact that $F_{\mathcal{H}}$ and $\rho_{\mathcal{H}}(\mathbf{R})$
have the analytic forms given in Eqs. \eqref{probability} and \eqref{f0}
will allow us to calculate easily the gradient of 
$\mathcal{F}_H[\mathcal{H}]$. Let us note that
from Eqs. \eqref{var-free-energy2} and \eqref{quantum-average} we observe that 
the free energy can be calculated simply as
\beq
\mathcal{F}_H[\mathcal{H}] = F_{\mathcal{H}} + 
      \int \mathrm{d}\mathbf{R} [ V(\mathbf{R}) -  \mathcal{V}(\mathbf{R})]
      \rho_{\mathcal{H}}(\mathbf{R}),
\label{var-free-energy3}
\eeq
where $V(\mathbf{R})$ is the BO energy of
ionic configuration $\mathbf{R}$ and $\mathcal{V}(\mathbf{R})$
is the trial harmonic energy for the same configuration.

%
%---------------------------
%
 
\subsection{The gradient of the free energy}
\label{gradient} 

Minimizing the free energy with respect to the trial 
harmonic Hamiltonian through a CG algorithm
requires the knowledge of the gradient of the 
free energy with respect to all the 
parameters in $\mathcal{H}$.
The trial $\mathcal{H}$ contains two group of 
parameters: the $\mathbf{R}_{{\rm eq}}$ equilibrium positions
and the $\Phi_{st}^{\alpha \beta}$ force-constants matrix.
Thus, the gradient of the free energy can be written as
$\boldsymbol{\nabla} \mathcal{F}_H[\mathcal{H}] =
(\boldsymbol{\nabla}_{\mathbf{R}_{\mathrm{eq}}}\mathcal{F}_H[\mathcal{H}], 
 \boldsymbol{\nabla}_{\Phi}\mathcal{F}_H[\mathcal{H}]) $,
where $\boldsymbol{\nabla}_{\mathbf{R_{\mathrm{eq}}}}\mathcal{F}_H[\mathcal{H}]$
is the gradient of the free energy with respect to the
equilibrium positions and 
$\boldsymbol{\nabla}_{\Phi}\mathcal{F}_H[\mathcal{H}]$
the gradient with respect to the force-constants matrix.

First of all, it can be shown that 
\beq
\boldsymbol{\nabla}_{\mathbf{R_{\mathrm{eq}}}}\mathcal{F}_H[\mathcal{H}] =
  -  \int \mathrm{d}\mathbf{R} [\mathbf{f}(\mathbf{R}) 
                          - \mathbf{f}_{\mathcal{H}}(\mathbf{R}) ]
                     \rho_{\mathcal{H}}(\mathbf{R}),
\label{gradient-eq}
\eeq
where $\mathbf{f}(\mathbf{R})$ is the vector formed
by all the atomic forces 
for the ionic configuration $\mathbf{R}$ and
$\mathbf{f}_{\mathcal{H}}(\mathbf{R})$
denotes the vector formed by the forces derived from $\mathcal{V}$. 
Note that the integral
with respect to the harmonic forces $\mathbf{f}_{\mathcal{H}}(\mathbf{R})$
vanishes, but, as we shall explain below, 
it is convenient to write the integral in
this form. On the other hand, the gradient with respect
to the force-constants matrix is given by 
\beqn
&& \boldsymbol{\nabla}_{\Phi} \mathcal{F}_H[\mathcal{H}] =   
   -\sum_{st\alpha \beta \mu} 
\sqrt{\frac{M_t}{M_s}} ( \epsilon_{\mu \mathcal{H}}^{s \alpha} 
    \boldsymbol{\nabla}_{\Phi} \ln a_{\mu \mathcal{H}}
   + \boldsymbol{\nabla}_{\Phi} \epsilon_{\mu \mathcal{H}}^{s \alpha} 
     ) \epsilon_{\mu \mathcal{H}}^{t \beta} \nonumber \\
&& \ \ \ \ \times  
  \int \mathrm{d}\mathbf{R} [f^{s \alpha}(\mathbf{R}) 
                          - f^{s \alpha}_{\mathcal{H}}(\mathbf{R}) ]
              (R^{t\beta} - R^{t\beta}_{\mathrm{eq}}) \rho_{\mathcal{H}}(\mathbf{R}).
\label{gradient-phi}
\eeqn 
The procedure to derive Eqs. \eqref{gradient-eq} and
\eqref{gradient-phi} is sketched in Appendix \ref{app-gradient}.
Let us note that both $\boldsymbol{\nabla}_{\Phi} a_{\mu \mathcal{H}}$
and $\boldsymbol{\nabla}_{\Phi} \epsilon_{\mu \mathcal{H}}^{s \alpha}$
are analytic functions of phonon frequencies and polarizations 
as shown in Appendix \ref{app-gradient}.

The CG minimization is started from a trial initial harmonic Hamiltonian
$\mathcal{H}_0$, which is defined following Eqs. \eqref{u-operator} 
and \eqref{harmonic-potential} from the $\mathbf{R}_{{\rm eq} 0}$ 
starting equilibrium positions 
and the starting $\Phi(0)$ force-constants matrix. 
After calculating the gradient as explained in 
Eqs. \eqref{gradient-eq} and
\eqref{gradient-phi}, the first CG step
allows us to update the equilibrium positions and
the force-constants matrix to $\mathbf{R}_{{\rm eq} 1}$
and $\Phi(1)$, from which we obtain the 
$\mathcal{H}_1$ Hamiltonian corresponding to first CG step.
Similarly, at each $j$ CG step of the minimization the
equilibrium positions and the force-constants matrix are updated
to $\mathbf{R}_{{\rm eq} j}$ and $\Phi(j)$, which define the 
Hamiltonian at step $j$, $\mathcal{H}_j$. The minimization
should be carried on until the gradient vanishes.    

%
%---------------------------
%

\subsection{Symmetries and the independent coefficients 
in the trial $\mathcal{H}$}
\label{parameters}

We consider that the anharmonic Hamiltonian given by the
SSCHA has the same symmetries as the harmonic Hamiltonian.
Therefore, at any CG step $j$ the Hamiltonian
$\mathcal{H}_j$ will respect the symmetries of the harmonic
Hamiltonian. Considering that $\mathcal{H}_j$
is determined by the $\Phi(j)$ force-constants matrix
and the $\mathbf{R}_{\mathrm{eq} j}$ equilibrium positions,
the symmetries of the Hamiltonian
are determined by the symmetries of both
$\mathbf{R}_{\mathrm{eq} j}$ and
$\Phi(j)$. In the SSCHA we consider translational, time-reversal
 and crystal symmetries
to determine the independent coefficients in the
equilibrium positions and the force-constants matrix.

If symmetries were neglected, throughout the minimization
the equilibrium positions could change in any direction
within the unit cell.
All these possible displacements can be described
with $3n$ size real vectors that form a vector space of dimension
$3n$,  where $n$ is the number of atoms in
the unit cell. The scalar product in this
vector space is defined as
\beq
\langle \boldsymbol{\chi},  
\boldsymbol{\xi} \rangle =
\sum_{\bar{s} \alpha}  \chi^{\bar{s} \alpha} \xi^{\bar{s} \alpha},
\label{scalar-prod-vec}
\eeq
where $\boldsymbol{\chi}$ and $\boldsymbol{\xi}$ are elements
of the vector space.
Let $\{ \boldsymbol{\chi}_{ \mathrm{(ns)}}(l) \}_{l=1,\dots,3n}$  
be an orthonormal basis of this vector space. 
Then, the vectors of the basis satisfy the
\beq
\langle \boldsymbol{\chi}_{ \mathrm{(ns)}} (l),  
\boldsymbol{\chi}_{ \mathrm{(ns)}} (l') \rangle =
 \delta_{ll'}
\label{orthonormality-vec}
\eeq
orthonormality condition. 
The bar in the atom index
$\bar{s}$ in Eq. \eqref{scalar-prod-vec} denotes 
that it is an atom of the unit cell 
and the (ns) subscript that the basis vectors have not been
symmetrized.
Thus, the equilibrium positions at a CG iteration $j$
could be given as
\beq
R_{{\rm eq} j}^{\bar{s} \alpha} = R_{{\rm eq} 0}^{\bar{s} \alpha} + \sum_{l=1}^{3n} 
                     \kappa_{j\mathrm{(ns)}}(l)
                     \chi^{\bar{s} \alpha}_{ \mathrm{(ns)}}(l),
\label{eqposfromkappa-ns}
\eeq
where the $\kappa_{j\mathrm{(ns)}}(l)$
coefficients would determine how much the atoms would be displaced
along $\boldsymbol{\chi}_{\mathrm{(ns)}}(l)$ at iteration $j$.
Obviously, $\kappa_{0\mathrm{(ns)}}(l) = 0$.
In the SSCHA however we allow the equilibrium
positions to change exclusively in the subspace
of this vector space that respects crystal symmetries.

In order to obtain a basis of the symmetrized subspace we need to take into account
all the $\hat{S} \equiv \{S,\mathbf{v}\}$ symmetry operations
of the space group of the crystal. Here, $S$ is a 3 $\times$ 3 
orthogonal matrix and $\mathbf{v}$ is the vector
defining the fractional translation of the crystal operation
$\hat{S}$. The $S$ matrices form the point group of the crystal.
We symmetrize the $\{ \boldsymbol{\chi}_{ \mathrm{(ns)}}(l) \}_{l=1,\dots,3n}$ 
basis vectors applying all the symmetry operations $\hat{S}$ 
as~\cite{0953-8984-21-39-395502}
\beq
\chi_{\mathrm{(s)}}^{\bar{s} \alpha}(l) = \frac{1}{N_S} \sum_{\hat{S}} 
               \sum_{\beta} S^{\alpha \beta}
               \chi_{\mathrm{(ns)}}^{\hat{S}^{-1}(\bar{s}) \beta}(l),
\label{symmetrize-vect-basis}                                       
\eeq
where $\hat{S}^{-1}$ is the
inverse symmetry operation of $\hat{S}$. 
The sum in Eq. \eqref{symmetrize-vect-basis}
runs over all the $N_S$ symmetry operations and
$\hat{S}^{-1}(\bar{s})$ labels the atom into which the $\bar{s}$-th atom
transforms after the application of $\hat{S}^{-1}$
modulo a lattice translation vector.
The (s) subscript denotes that the vectors 
respect symmetries.
Note that Eq. \eqref{symmetrize-vect-basis} is commonly used in 
DFT codes to symmetrize the forces on the atoms, and when the 
electronic $\mathbf{k}$-point mesh is reduced by symmetry.
When we symmetrize the basis vectors as shown in Eq. \eqref{symmetrize-vect-basis},
many of these $\boldsymbol{\chi}_{\mathrm{(s)}}(l)$ vectors become
linearly dependent. We pick exclusively the linearly
independent vectors applying a Gram-Schmidt orthonormalization
procedure. The basis vectors of the symmetrized subspace are labeled as
$\{ \boldsymbol{\chi}(l) \}_{l=1,\dots,n_w}$, where $n_w$ is the
number of linearly independent basis vectors after
the symmetrization.
The  Gram-Schmidt orthonormalization guarantees that the
$\boldsymbol{\chi}(l)$ vectors satisfy Eq. \eqref{orthonormality-vec}.
The value of $n_w$ must be equal to the number of
free internal parameters in the
Wyckoff positions of the crystal structure.

Once we have determined the symmetry reduced basis, 
at a given iteration $j$ in the CG minimization the equilibrium position
of the atoms can be described as
\beq
R_{{\rm eq} j}^{\bar{s} \alpha} = R_{{\rm eq} 0}^{\bar{s} \alpha} + 
                     \sum_{l=1}^{n_w} \kappa_j(l)
                     \chi^{\bar{s} \alpha}(l),
\label{eqposfromkappa}
\eeq
where the $\kappa_j(l)$ coefficient
determines how much the atoms are displaced along 
the symmetrized $\boldsymbol{\chi}(l)$ direction
at iteration $j$.  
Then, it is easy to relate 
$\boldsymbol{\nabla}_{\mathbf{R_{\mathrm{eq}}}}\mathcal{F}_H[\mathcal{H}_j]$ 
with the derivatives of the free energy with respect to the
$\kappa_j(l)$ coefficients introduced in Eq.
\eqref{eqposfromkappa}:
\beq
\frac{\partial \mathcal{F}_H[\mathcal{H}_j]}{\partial \kappa_j(l)} =
\sum_{s \alpha} \chi^{s \alpha}(l) 
\frac{\partial \mathcal{F}_H[\mathcal{H}_j]}{\partial R_{{\rm eq} j}^{s \alpha}},
\label{grad-coeff-pos}
\eeq
where $\frac{\partial \mathcal{F}_H[\mathcal{H}_j]}{\partial R_{{\rm eq} j}^{s \alpha}}$
is given in Eq. \eqref{gradient-eq}.

In order to determine how $\Phi$ can change in the
SSCHA minimization respecting crystal, time-reversal 
and translational symmetries, we proceed in an analogous way.
In general $\Phi$ is a matrix that belongs to the
group of $3N \times 3N$ Hermitian matrices. 
The Hermitian matrices form a vector space
and the scalar product between two elements
of the vector space is defined as
\beq
\langle \mathcal{G}, \mathcal{T} \rangle =
\sum_{s t \alpha \beta}   \mathcal{G}_{st}^{\alpha \beta}   
                          \mathcal{T}_{st}^{\alpha \beta *} ,
\label{scalar-prod-mat}
\eeq
where $\mathcal{G}$ and $\mathcal{T}$ are
two elements of the vector space.
We start with the subspace of this vector space that
preserves translational symmetries but has not been
symmetrized with the $\hat{S}$ crystal symmetry operations
nor time-reversal. Let 
$\{\mathcal{G}_{\mathrm{(ns)}}(m) 
\}_{m=1,\dots,(3n)^2 N_1 N_2 N_3}$ be an orthonormal
basis of this vector space so that
\beq
\langle \mathcal{G}_{\mathrm{(ns)}} (m), \mathcal{G}_{\mathrm{(ns)}} (m') \rangle =
   \delta_{mm'}.
\label{orthonormality-mat}
\eeq
Thanks to Bloch's theorem, 
the dimension of this vector space is $(3n)^2 \times N_1 \times N_2\times N_3$,
where $N_1 \times N_2 \times N_3$ is the supercell size.
As any matrix belonging to this vector space respects translational
symmetries, the Fourier transform of a matrix described in the
$\{\mathcal{G}_{\mathrm{(ns)}}(m) \}_{m=1,\dots,(3n)^2 N_1 N_2 N_3}$  
basis is block-diagonal and can be defined with a single 
$\mathbf{q}$ vector in the first Brillouin zone (1BZ).
Thus, if only transitional symmetries were considered,
the evolution of the force-constants matrix in the minimization
could be described as
\beq
\Phi (j) = \sum_{m=1}^{(3n)^2N_1  N_2 N_3} c_{j \mathrm{(ns)}}(m) 
           \mathcal{G}_{\mathrm{(ns)}} (m),
\label{decompose-fc-stepj-ns}
\eeq
where the $c_{j \mathrm{(ns)}}(m)$ coefficients would determine
the value of the force-constants matrix at CG step $j$.
Nevertheless, in the SSCHA we allow the force-constants
matrix to vary exclusively in the subspace of this vector 
space that respects the $\hat{S}$ crystal symmetries~\cite{RevModPhys.40.1} 
and time-reversal symmetry. 

The elements of the basis are symmetrized according to
the $\hat{S}$ symmetry operations and time-reversal as
shown in Appendix \ref{app-symmetry}. After the
symmetrization, the basis is reduced making use of a Gram-Schmidt orthonormalization
procedure so that only the linearly independent elements of the
symmetrized basis are considered.
This process yields a new $\{\mathcal{G}(m)\}_{m=1,\dots,N_R}$
basis that respects translational, crystal and
time-reversal symmetries.
$N_R$ is the dimension of the fully symmetrized subspace.
We construct the force-constants matrix as  
\beq
\Phi(j) = \sum_{m=1}^{N_R} c_j(m) 
           \mathcal{G}_{st}^{\alpha \beta} (m).
\label{decompose-fc-stepj}
\eeq
The $c_j(m)$ coefficients unambiguously determine the
force-constants matrix at each CG iteration $j$.
With the $\mathcal{G}(m)$ matrices it is
easy to relate the derivative of the free energy with respect to
the $c_j(m)$ coefficients with
$\boldsymbol{\nabla}_{\Phi} \mathcal{F}_H[\mathcal{H}_j]$.
From Eq. \eqref{decompose-fc-stepj} straightforwardly
\beq
\frac{\partial \mathcal{F}_H[\mathcal{H}_j]}{\partial c_j(m)} =
\sum_{st \alpha \beta}  \mathcal{G}_{st}^{\alpha \beta}(m)
\frac{\partial \mathcal{F}_H[\mathcal{H}_j]}{\partial \Phi_{st}^{\alpha \beta}(j)},
\label{grad-coeff-phi}
\eeq
where the 
$\frac{\partial \mathcal{F}_H[\mathcal{H}_j]}{\partial \Phi_{st}^{\alpha \beta}(j)}$
derivatives are given in Eq. \eqref{gradient-phi}.

In order to illustrate the reduction of coefficients, let us
consider a $4 \times 4 \times 4$ supercell of a rock-salt
structure. In this case, the \emph{a priori} 2304 $c_{\mathrm{(ns)}}(m)$
free parameters in the force-constants matrix in Eq. \eqref{decompose-fc-stepj-ns}
are reduced to simply 50 $c(m)$ parameters in Eq. \eqref{decompose-fc-stepj}.

Considering the independent coefficients that we have
found after the symmetry analysis, 
we can write the gradient of the free energy as
$\boldsymbol{\nabla} \mathcal{F}_H[\mathcal{H}] =
(\boldsymbol{\nabla}_{\kappa}\mathcal{F}_H[\mathcal{H}], 
 \boldsymbol{\nabla}_{c}\mathcal{F}_H[\mathcal{H}]) $. The
number of components in this gradient is much fewer than the 
components in 
$(\boldsymbol{\nabla}_{\mathbf{R}_{\mathrm{eq}}}\mathcal{F}_H[\mathcal{H}], 
 \boldsymbol{\nabla}_{\Phi}\mathcal{F}_H[\mathcal{H}]) $.
Therefore, in the SSCHA we work with $\boldsymbol{\nabla} \mathcal{F}_H[\mathcal{H}] =
(\boldsymbol{\nabla}_{\kappa}\mathcal{F}_H[\mathcal{H}], 
 \boldsymbol{\nabla}_{c}\mathcal{F}_H[\mathcal{H}]) $.
At a given iteration $j$ of the CG minimization, the
components of $\boldsymbol{\nabla}_{\kappa}\mathcal{F}_H[\mathcal{H}_j]$
are given in Eq. \eqref{grad-coeff-pos} and the components of 
$\boldsymbol{\nabla}_{c}\mathcal{F}_H[\mathcal{H}_j]$ in
Eq. \eqref{grad-coeff-phi}. 

%%%%%%%%%%%%%
% THE SSCHA %             
%%%%%%%%%%%%%

\section{The stochastic implementation of the self-consistent harmonic
            approximation}
\label{sscha}

The calculation of the integrals in Eqs. 
\eqref{var-free-energy3}-\eqref{gradient-phi} needed to get the free energy
and its gradient is a complicated task~\cite{PhysRev.128.2589}.
In principle, it requires the calculation of high-order
$\phi_{s_1 \dots s_n}^{\alpha_1 \dots \alpha_n}$ 
anharmonic coefficients that allow an accurate estimation of
the $\mathbf{f}$ forces and the $V$ potential.
Third order anharmonic coefficients can be calculated
nowadays through first-principles calculations using the $2n+1$ 
theorem~\cite{PhysRevB.87.214303}. However, 
one should note that third order terms 
do not contribute in Eq. \eqref{gradient-phi}.
The reason is that the integrand is odd for third order terms
and, thus, the integral vanishes by symmetry. 
Therefore, one needs to go at least to
the fourth order to apply the SCHA. The calculation of the fourth order anharmonic
coefficients is extremely cumbersome as it requires performing
first order numerical derivatives of third order anharmonic terms
or second order numerical derivatives of dynamical matrices
calculated in supercells~\cite{PhysRevB.59.6182,PhysRevB.68.220509,PhysRevB.82.104504,
PhysRevLett.106.165501,errea:112604,PhysRevLett.99.176802}. 
Consequently, calculating
fourth-order anharmonic coefficients remains a complicated 
computational problem and these coefficients have been calculated 
\abinitio{} exclusively
for some specific $\mathbf{q}$ points in the 1BZ
or in very simple
crystal structures~\cite{PhysRevB.68.220509,PhysRevB.82.104504,
PhysRevLett.106.165501,errea:112604,PhysRevB.59.6182,PhysRevLett.99.176802}.   
Therefore, the SCHA has been applied calculating explicitly 
the fourth-order anharmonic coefficients in the whole 1BZ 
purely \abinitio{} only in the high-pressure simple cubic
phase of calcium~\cite{PhysRevLett.106.165501,errea:112604}.
Moreover, the restriction to fourth-order terms is an approximation
that could be inappropriate and should be verified case by case.

In the SSCHA we take a different approach and, instead of
calculating $\phi_{i_1 \dots i_n}^{\alpha_1 \dots \alpha_n}$ 
coefficients, we evaluate the integrals stochastically 
using suitably chosen ionic configurations in supercells
without assuming any Taylor development. 
The stochastic evaluation of the quantum statistical
average of any observable is performed taking advantage of the 
analytic behavior of $\rho_{\mathcal{H}}(\mathbf{R})$ 
and making use of importance
sampling and reweighting techniques.
  
%
%---------------------------
%
 
\subsection{Stochastic calculation of the gradient}
\label{sotchastic-gradient} 

As it was mentioned above, the minimization of 
$\mathcal{F}_H[\mathcal{H}]$ is started
from an arbitrary harmonic Hamiltonian $\mathcal{H}_0$.  
Then, we create a set of $\{\mathbf{R}_I\}_{I=1,\dots,N_c}$ ionic 
configurations in the supercell 
according to the $\rho_{\mathcal{H}_0}(\mathbf{R})$ 
distribution given in Eq. \eqref{probability}.
The distribution is determined by the starting
$\mathbf{R}_{{\rm eq} 0}$ equilibrium positions
and the starting $\Phi(0)$ force-constants matrix,
and can be created using random numbers
generated with a pure Gaussian distribution as shown
in Appendix \ref{app-stat}.
According to the importance sampling technique,
any quantum statistical average of an operator
that exclusively depends on the atomic positions
can be evaluated as an average of the operator 
over the created $N_c$ configurations. Namely,
\beq
\int \mathrm{d}\mathbf{R} O(\mathbf{R}) \rho_{\mathcal{H}_0}(\mathbf{R}) \simeq 
           \frac{1}{N_c} \sum_{I=1}^{N_c} O(\mathbf{R}_I) \equiv \langle O \rangle,
\label{quantum-average-stat}
\eeq
where $O(\mathbf{R}_I)$ denotes the value of the operator 
$O(\mathbf{R})$ at
the configuration $\mathbf{R}_I$. 
In Eq. \eqref{quantum-average-stat}
the equality holds when $N_c \to \infty$ and the error
in the stochastic evaluation vanishes.
Therefore, we evaluate BO energies and atomic forces
in the $\{\mathbf{R}_I\}_{I=1,\dots,N_c}$ configurations,
$V(\mathbf{R}_I)$ and $\mathbf{f}(\mathbf{R}_I)$,
respectively, and calculate the integrals
in Eqs. \eqref{var-free-energy3}-\eqref{gradient-phi}
following the stochastic procedure of Eq. \eqref{quantum-average-stat}.
Once these are computed,  
$\boldsymbol{\nabla}\mathcal{F}_H[\mathcal{H}_0]$
can be obtained stochastically and the first CG
step can be performed to obtain $\mathcal{H}_1$. 

After the first CG step, the ionic configurations 
should in principle be regenerated as in Eq. \eqref{probability}
using the new trial Hamiltonian $\mathcal{H}_1$, 
which is defined by the $\mathbf{R}_{{\rm eq} 1}$ equilibrium positions
and the $\Phi(1)$ force-constants matrix.
Thus, in order to calculate the gradient we 
should recalculate BO energies and atomic forces in the supercell in the 
new set of configurations defined by $\rho_{\mathcal{H}_1}(\mathbf{R})$. 
Considering that in general hundreds of CG steps are needed to
find the minimum of the free energy, calculating BO energies and forces
from first-principles at each CG step would make the method prohibitively time-demanding. 
We adopt a reweighting procedure to avoid this issue and use the 
BO energies and atomic forces of the initial 
$\{\mathbf{R}_I\}_{I=1,\dots,N_c}$ set throughout the CG minimization.
At step $j$ of the CG minimization, this is achieved including 
the $\rho_{\mathcal{H}_j}(\mathbf{R}) / \rho_{\mathcal{H}_0}(\mathbf{R})$ 
factor in the importance sampling evaluation of the integrals.
Note that in the first $j=0$ step the factor is equal to one.
Therefore, at step $j$ of the CG minimization
the integral in  Eq. \eqref{quantum-average-stat}
is computed as if the initial $\{\mathbf{R}_I\}_{I=1,\dots,N_c}$
set was generated according to 
$\rho_{\mathcal{H}_j}(\mathbf{R})$, namely
\beqn
\int \mathrm{d}\mathbf{R} O(\mathbf{R}) \rho_{\mathcal{H}_j}(\mathbf{R}) 
  & \simeq & 
           \frac{1}{N_c} \sum_{I=1}^{N_c} O(\mathbf{R}_I) 
          \frac{\rho_{\mathcal{H}_j}(\mathbf{R}_I)}{
               \rho_{\mathcal{H}_0}(\mathbf{R}_I)} \nonumber \\
   & = & \langle O \rho_{\mathcal{H}_j} / \rho_{\mathcal{H}_0} \rangle.
\label{quantum-average-stat-j}
\eeqn
The stochastic error in Eq. \eqref{quantum-average-stat-j} 
can be evaluated as
\beq
\Delta \langle O \rho_{\mathcal{H}_j} / \rho_{\mathcal{H}_0} \rangle
= \frac{1}{\sqrt{N_c}} \sqrt{s^2_{ O \rho_{\mathcal{H}_j} / \rho_{\mathcal{H}_0}}},
\label{error-stat-j}
\eeq
where 
\beq
s^2_{P} = \frac{1}{N_c - 1} \sum_{I=1}^{N_c}   
   [P(\mathbf{R}_I) - \langle P \rangle ]^2
\label{variance}
\eeq
is the variance of function $P(\mathbf{R})$\footnote{
At step $j$ of the CG minimization, one could 
also calculate the integral in Eq. \eqref{quantum-average-stat}
as $\int \mathrm{d}\mathbf{R} O(\mathbf{R}) \rho_{\mathcal{H}_j}(\mathbf{R}) 
\simeq 
   \frac{ \langle O \rho_{\mathcal{H}_j}  / \rho_{\mathcal{H}_0} \rangle }{
    \langle \rho_{\mathcal{H}_j} / \rho_{\mathcal{H}_0} \rangle}$. 
In this case, the error would not be given by Eq. \eqref{error-stat-j},
but a slightly modified version. Both ways of calculating the integral
become identical in the $N_c \to \infty$ limit.
}. Following Eq. \eqref{quantum-average-stat-j}, the 
free energy in Eq. \eqref{var-free-energy3} and its gradient
in Eqs. \eqref{gradient-eq} and \eqref{gradient-phi} are
calculated at a given iteration $j$ of the CG minimization
simply as
\beqn
\mathcal{F}_H[\mathcal{H}_j]  \simeq  F_{\mathcal{H}_j} + 
     \frac{1}{N_c} \sum_{I=1}^{N_c} [ V(\mathbf{R}_I) -  \mathcal{V}_j(\mathbf{R}_I)]
     \frac{\rho_{\mathcal{H}_j}(\mathbf{R}_I)}{
               \rho_{\mathcal{H}_0}(\mathbf{R}_I)} \nonumber \\ 
\label{potential-stat} \\
\boldsymbol{\nabla}_{\mathbf{R_{\mathrm{eq}}}}\mathcal{F}_H[\mathcal{H}_j]
    \simeq   - 
  \frac{1}{N_c} \sum_{I=1}^{N_c} 
    [\mathbf{f}(\mathbf{R}_I) 
                          - \mathbf{f}_{\mathcal{H}_j}(\mathbf{R}_I) ]
     \frac{\rho_{\mathcal{H}_j}(\mathbf{R}_I)}{
               \rho_{\mathcal{H}_0}(\mathbf{R}_I)} \nonumber \\ 
 \label{grad-eq-estat} 
\eeqn
\begin{widetext}
\beqn
\boldsymbol{\nabla}_{\Phi} \mathcal{F}_H[\mathcal{H}_j]  \simeq    
   -\sum_{st\alpha \beta \mu} 
\sqrt{\frac{M_t}{M_s}} ( \epsilon_{\mu \mathcal{H}_j}^{s \alpha} 
    \boldsymbol{\nabla}_{\Phi} \ln a_{\mu \mathcal{H}_j}
   + \boldsymbol{\nabla}_{\Phi} \epsilon_{\mu \mathcal{H}_j}^{s \alpha} 
     ) \epsilon_{\mu \mathcal{H}_j}^{t \beta}   \frac{1}{N_c} \sum_{I=1}^{N_c}  
    [f^{s \alpha}(\mathbf{R}_I) 
                          - f^{s \alpha}_{\mathcal{H}_j}(\mathbf{R}_I) ]
              (R^{t\beta}_I - R^{t\beta}_{\mathrm{eq} j})
     \frac{\rho_{\mathcal{H}_j}(\mathbf{R}_I)}{
               \rho_{\mathcal{H}_0}(\mathbf{R}_I)}. \nonumber \\
\label{grad-phi-estat} 
\eeqn
\end{widetext}
In Eqs. \eqref{potential-stat}-\eqref{grad-phi-estat} the equality
holds when $N_c \to \infty$.

Let us note that, despite the contribution of 
$\mathbf{f}_{\mathcal{H}}$ in
Eq. \eqref{grad-eq-estat}  vanishes and is 
analytic in Eq. \eqref{grad-phi-estat} (see Appendix \ref{app-gradient}), it is 
convenient to keep this contribution explicitly in the
stochastic evaluation of the gradient. 
Similarly, in the stochastic evaluation of the free energy in
Eq. \eqref{potential-stat}, it is convenient to
keep the $\mathcal{V}$ contribution even if it
is analytic as well. The reason is that 
in this way the stochastic analysis
is performed exclusively on the anharmonic part of the forces 
or the BO energies, reducing the stochastic error.
Therefore, at each step $j$ of the CG
minimization $\mathbf{f}_{\mathcal{H}_j}(\mathbf{R}_I)$ and
$\mathcal{V}_j(\mathbf{R}_I)$ are calculated,
which are analytic functions of the $\omega_{\mu \mathcal{H}_j}$
frequencies and $\epsilon_{\mu \mathcal{H}_j}^{s \alpha}$
polarizations.
The fact that including $\mathbf{f}_{\mathcal{H}_j}(\mathbf{R}_I)$ and
$\mathcal{V}_j(\mathbf{R}_I)$ in the evaluation of
the free energy and its gradient is beneficial for the
stochastic approach is exemplified if we assume the $V(\mathbf{R})$
potential is perfectly harmonic and the initial  
$\mathcal{H}_0$ is the harmonic Hamiltonian. 
Then, the gradient obtained in the first step stochastically
is exactly zero, with no stochastic error, as
$\mathbf{f}(\mathbf{R}_I) - \mathbf{f}_{\mathcal{H}_0}(\mathbf{R}_I)
= 0$. Similarly, the free energy would not have any 
stochastic error since  
$V(\mathbf{R}_I) - \mathcal{V}_0(\mathbf{R}_I)
= 0$, and  $\mathcal{F}_H[\mathcal{H}_0] = F_{\mathcal{H}_0}$.
If $\mathbf{f}_{\mathcal{H}_0}(\mathbf{R}_I)$ and
$\mathcal{V}_0(\mathbf{R}_I)$ were not included in the sctochastic
evaluation of the integrals by using their analytic expression 
instead, the stochastic error would not vanish. 

As shown above, the free energy and its gradient 
can be obtained calculating BO energies and ionic forces on
supercells with suitably chosen ionic configurations. 
The calculation of the BO energies and forces can be performed
using a model or \abinitio{} potentials.
It is noteworthy that calculating the forces from first-principles
requires a negligible additional effort in a total energy calculation
because of the Hellmann-Feynman theorem~\cite{hellmann,PhysRev.56.340}.  
Hence, the way of minimizing the gradient sketched above is very
convenient for applying the SSCHA fully from first-principles,
specially, due to the reweighting procedure.

%
%---------------------------
%
 
\subsection{Stopping criteria and calculation flowchart}
\label{flowchart}  
 
\begin{figure}[t]
\includegraphics[width=0.50\textwidth]{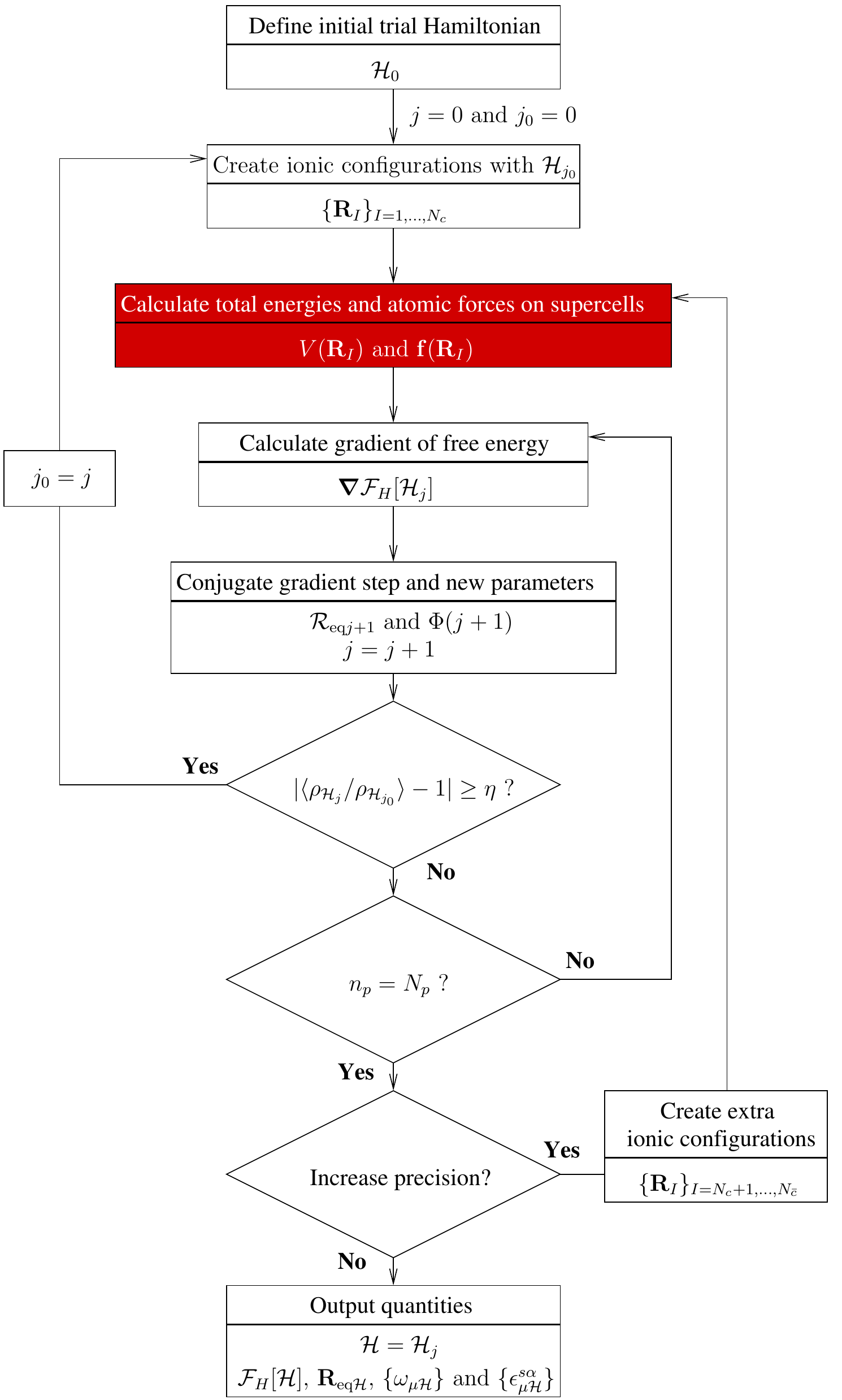}
\caption{(Color online) Schematic representation of the SSCHA calculation
flowchart. See Sec. \ref{flowchart} of the main text 
for the definition of the symbols. The step in the 
algorithm marked in red, the calculation of the
total energies and forces on supercells, is performed
with any external total-energy-force engine. In case
an \abinitio{} approach is taken, it represents
all the computer time of the SSCHA minization.}
\label{flowchart-fig}
\end{figure}

In principle, the minimization should continue till all the 
components of the gradient are smaller than a given threshold 
value. The threshold value should be chosen so that phonon
frequencies and equilibrium positions are converged
with respect to it. When all components of the
gradient are smaller than the threshold value the minimum
of the free energy has been found and 
$\mathcal{H}_j$ is the harmonic Hamiltonian
that minimizes it, $j$ being the final step. 
Then, the $\mathbf{R}_{{\rm eq} j}$  
are the final equilibrium atomic positions, $\{ \omega_{\mu \mathcal{H}_j} \}$
form the SSCHA phonon spectra and $\{ \epsilon_{\mu \mathcal{H}_j}^{s \alpha} \}$
are the final polarization vectors, which are obtained
diagonalizing the $\Phi(j)$ force-constants matrix,
and $\mathcal{F}_H[\mathcal{H}_j]$ is the final 
vibrational free energy.

Considering that
$ \boldsymbol{\nabla}_{\kappa} \mathcal{F}_H[\mathcal{H}_j]$ and 
$ \boldsymbol{\nabla}_{c} \mathcal{F}_H[\mathcal{H}_j]$ have different 
units, we should use different threshold values for each of them,
let's say $\zeta_{\kappa}$ and $\zeta_{c}$, respectively.
Thus, if
\beq
|\nabla^i_h \mathcal{F}_H[\mathcal{H}_j]| < \zeta_h, 
\label{conv-gradient}
\eeq
where $h$ denotes $\kappa$ or $c$ 
and $\nabla^i_h \mathcal{F}_H[\mathcal{H}_j]$ is the
$i$-th component of the gradient at CG step $j$,
the $i$-th parameter is not updated in the $j+1$ step.

Due to the stochastic origin of the method, we can 
account for the error of the gradient as shown in Eq. \eqref{error-stat-j} 
in the minimization 
and devise a second stopping criteria.
We define 
a meaningfulness parameter $\theta$ and in case
\beq
\theta \Delta ( \nabla^i \mathcal{F}_H[\mathcal{H}_j]) 
> | \nabla^i \mathcal{F}_H[\mathcal{H}_j] |,
\label{meaningfulness}
\eeq 
where $\Delta ( \nabla^i \mathcal{F}_H[\mathcal{H}_j])$ is
the stochastic error of the $i$-th component of the 
gradient, the $i$-th parameter is not updated in the $j+1$ step.
From now on, let $n_p$ be the number of components of the gradient
satisfying  Eq. \eqref{conv-gradient} or 
\eqref{meaningfulness},
and $N_p$ the total number of components of the 
gradient.

Moreover, if
$\langle \rho_{\mathcal{H}_j} / \rho_{\mathcal{H}_0} \rangle$
deviates significantly from 1 the gradient cannot be stochastically
evaluated accurately because the initial set of configurations
does not represent closely the $\rho_{\mathcal{H}_j}(\mathbf{R})$
distribution. Hence, if 
\beq
| \langle \rho_{\mathcal{H}_j} / \rho_{\mathcal{H}_0} \rangle - 1 | \geq \eta,
\label{poor-stat}
\eeq 
where $\eta$ is a small positive number, generally between 0.2 and 0.3,
we consider that the stochastic
evaluation of the gradient is poor and the minimization is stopped 
at the CG step $j$. 

The SSCHA calculation flowchart is sketched in Fig. 
\ref{flowchart-fig}. As mentioned above, 
the minimization is stopped at step $j$ if (i)
$n_p = N_p$ or (ii) the stochastic evaluation
is poor according to Eq. \eqref{poor-stat}. If condition (ii) is satisfied
at CG step $j$, the $\rho_{\mathcal{H}_j}(\mathbf{R})$
probability distribution is used to create a new set 
of $\{ \mathbf{R}_I \}_{I=1,\dots,N_c}$ configurations
for which total energies and atomic forces
are recomputed. Then, the minimization continues using this new
set of configurations. On the other hand, if 
condition (i) is fulfilled, one needs to see whether 
the stochastic accuracy is satisfactory.
If it is so, the calculation is finished and the minimum is found.
If, on the contrary, one wants to increase 
the precision in the evaluation of the gradient,
the number of configurations should be increased in order to 
reduce the error. This can be done 
generating $N_{\bar{c}} - N_c$ new configurations with the
last Hamiltonian used to generate configurations in order
to increase the size of the set to $N_{\bar{c}}$. Then,
new total energies and atomic forces are calculated in the new 
$N_{\bar{c}} - N_c$ configurations and the process continues till the 
stochastic uncertainty is reduced up to a satisfactory level.
This should be noted in the convergence of the phonon spectra.

The calculation of total energies and atomic forces needed
in the SSCHA algorithm can be performed at any degree of theory
with any external total-energy-force engine. If an \abinitio{}
approach is taken, practically all the computer time goes
in the calculation of the total energies and forces.
The algorithm sketched in Fig. \ref{flowchart-fig}
is devised to minimize the number of calls to the
total-energy-force engine. For instance, one
can start with a small number of $N_c$ until 
the calculation stops because $n_p = N_p$. The
size of the set can be increased at this point to 
gain accuracy. Thus, the number of calls to the 
total-energy-force engine can be effectively optimized
in the SSCHA algorithm.
Obviously, the number of
total energy and atomic force calculations is reduced in case 
the starting $\mathcal{H}_0$ is close to the  
$\mathcal{H}$ Hamiltonian that minimizes the free energy.
This way, the need to redefine a new set of of configurations
might be avoided. Anyway, the final result is independent of the
starting $\mathcal{H}_0$ .

%
%---------------------------
%
 
\subsection{Temperature dependence in the SSCHA}
\label{temperature-sec}

The temperature dependence in the SSCHA is naturally incorporated.
As shown in Eqs. \eqref{var-free-energy3}-\eqref{gradient-phi},
the free energy and its gradient depend on temperature 
through the temperature dependence of $F_{\mathcal{H}}$
(see Eq. \eqref{f0}) and $\rho_{\mathcal{H}}(\mathbf{R})$,
which depends on temperature via the normal lengths  
$a_{\mu \mathcal{H}}$ (see Eqs. \eqref{probability} and 
\eqref{normal-length}). When creating the set of configurations
$\{\mathbf{R}_I\}_{I=1,\dots,N_c}$ the temperature
dependence is incorporated as the $a_{\mu \mathcal{H}}$
normal lengths are used to generate the set as noted in
Appendix \ref{app-stat}.
Thus, in principle, one should use a given set of configurations
for each temperature, calculating new forces and BO energies 
for each temperature. 

Nevertheless, a recycling scheme can be adopted to use
the set of configurations created with a given temperature
$T_0$ (and the forces and BO energies calculated for them) 
to perform the minimization at a different temperature
$T$. In order to do so, in Eqs. 
\eqref{potential-stat}-\eqref{grad-phi-estat} we modify the factor used in the
reweighting as 
\beq
\frac{\rho_{\mathcal{H}_j}(\mathbf{R}_I)}{
               \rho_{\mathcal{H}_0}(\mathbf{R}_I)}
\to
\frac{\rho_{\mathcal{H}_j}(\mathbf{R}_I,T)}{
               \rho_{\mathcal{H}_0}(\mathbf{R}_I,T_0)},
\label{recycling-t}
\eeq
where $\rho_{\mathcal{H}_0}(\mathbf{R}_I,T_0)$ is the 
probability distribution function used to generate
the $\{\mathbf{R}_I\}_{I=1,\dots,N_c}$ configurations
and $\rho_{\mathcal{H}_j}(\mathbf{R}_I,T)$ 
is the probability distribution at CG step $j$ for
temperature $T$ at which the free energy wants to be minimized.
We should note that when adopting this recycling scheme
at $j=0$ 
$\langle \rho_{\mathcal{H}_0}(T) / \rho_{\mathcal{H}_0} (T_0) \rangle$
is different from one. 
Thus, we use the criteria defined in Eq. \eqref{poor-stat}
to discern whether recycling the
configurations created with a different temperature  
is valid.

%%%%%%%%%%%%%%%%%%%%%%
% APPLICATION TO PtH %
%%%%%%%%%%%%%%%%%%%%%%

\section{Application of the stochastic self-consistent harmonic 
approximation to platinum hydride at high pressure}
\label{pth}

Motivated by the quest
for metallic and superconducting hydrogen
at very high pressure~\cite{PhysRevLett.21.1748},
many first-principles calculations based on the 
harmonic approximation have been performed
in the last years in compressed hydrides predicting
high values for the superconducting critical 
temperature ($T_c$)~\cite{Kim16022010,
PhysRevLett.107.117002,Gao26012010,gao:107002,
cudazzo:257001,martinez-canales:087005}.
Nevertheless, the only experimental evidence so far of superconductivity
in hydrides at high pressure was found in SiH$_4$ around 100 GPa with
$T_c = 17$ K~\cite{Eremets14032008}. However, 
it is not clear whether the superconductivity
of SiH$_4$ was actually measured. Degtyareva \etal{} proposed
that  PtH could have been formed in 
that experiment if silane decomposed
releasing hydrogen that reacted with the platinum 
electrodes~\cite{Degtyareva20091583}.
It has been 
argued~\cite{Degtyareva20091583,PhysRevB.84.054543}
that the formation of PtH might explain the x-ray diffraction
pattern observed in Ref.~\onlinecite{Eremets14032008},
and that the observed superconductivity might be attributed
to the superconductivity of PtH as its calculated
$T_c$ is not far from the measured 
value~\cite{PhysRevB.83.214106,
PhysRevB.84.054543,PhysRevLett.107.117002}.
However, in PtH $T_c$ has been calculated within the harmonic approximation
and it is not clear whether anharmonicity might affect this result,
as it does in the very similar PdH~\cite{PhysRevLett.111.177002}.
We apply the SSCHA method to high-pressure PtH
to shine light on this issue.

\subsection{Calculation details}
\label{technical-pth}

\begin{figure}[t!]
\includegraphics[width=0.49\textwidth]{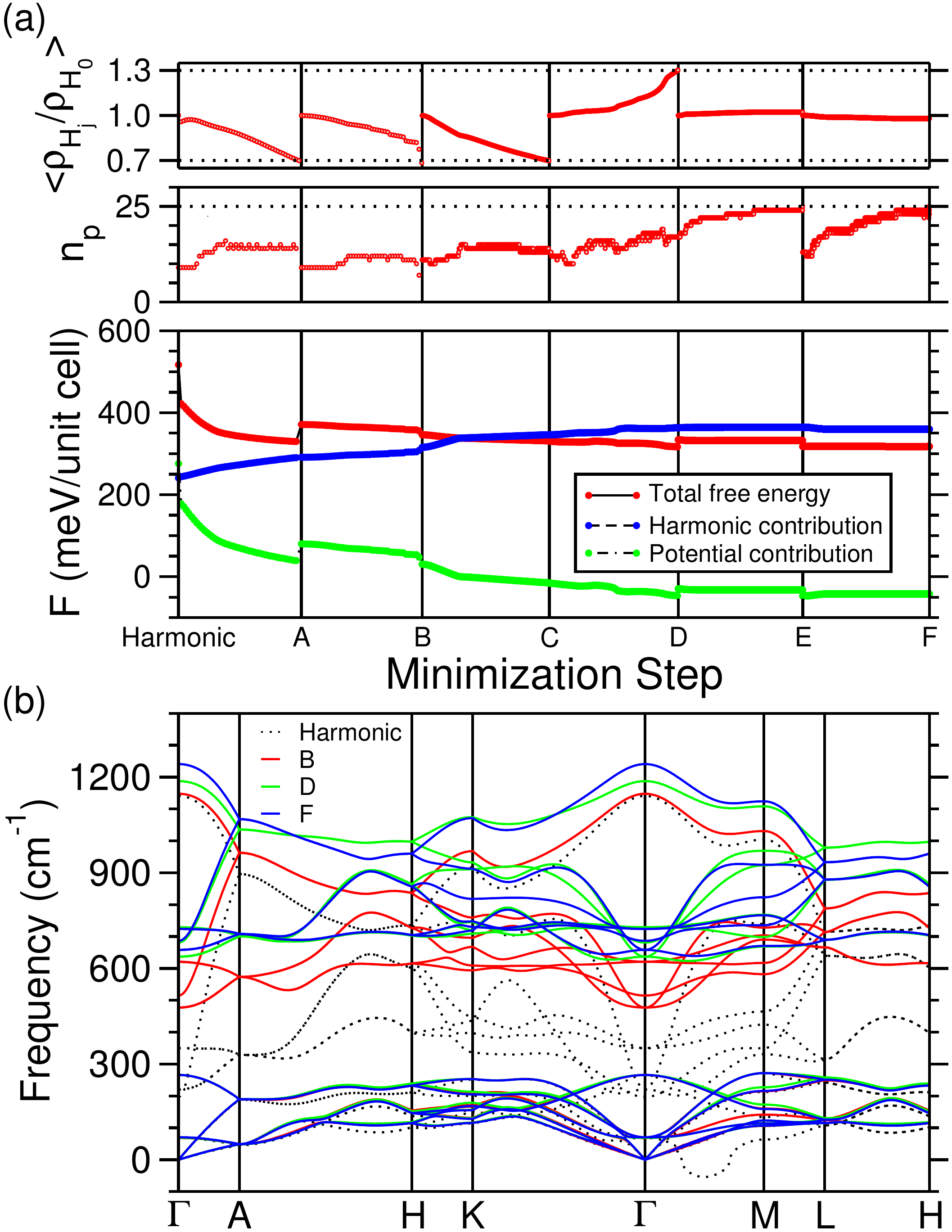}
\caption{ (Color online) The evolution of the SSCHA calculation
in PtH at 0 K and 100 GPa. 
The starting Hamiltonian is the harmonic one. 
The parameters of the calculation
are $\theta = 1$, $\zeta_{c} = 0.0003$ a.u.$^2$ and $\eta = 0.3$
(see Sec. \ref{technical-pth}).
In (a) we depict the evolution of 
$\langle \rho_{\mathcal{H}_j} / \rho_{\mathcal{H}_0} \rangle$, 
$n_p$, the number of parameters that 
are not updated according to Eqs. \eqref{conv-gradient} or \eqref{meaningfulness}, 
and the free energy throughout the minimization. 
In this calculation the total number of parameters in the gradient is $N_p = 25$.
In these figures each point represents a CG step.
The total vibrational free energy
$\mathcal{F}_H[\mathcal{H}_j] = F_{\mathcal{H}_j} + \mathrm{tr} [ \rho_{\mathcal{H}_j}
(V - \mathcal{V}_j) ]$, the harmonic contribution
$F_{\mathcal{H}_j}$, and the potential contribution $\mathrm{tr} [ \rho_{\mathcal{H}_j}
( V - \mathcal{V}_j) ]$ are specified in the bottom panel. 
In (b) the evolution of the phonon spectra is plotted presenting
the results at iterations B, D, and F, together with the starting
harmonic phonon spectra.}
\label{example-fig}
\end{figure}

We apply the SSCHA in PtH at 100 GPa and 0 K. Even if at
ambient conditions Pt and H are immiscible, at high
pressure platinum hydride can be synthesized~\cite{PhysRevB.83.214106}.
Experimentally it was observed that above 42 GPa PtH adopts
a hexagonal closed-packed (hcp) structure~\cite{PhysRevB.83.214106}
and, consequently, we use the hcp phase in our calculations.
We take the lattice parameters that minimize the electronic
energy at 100 GPa, namely $a=5.1203$ a.u. and $c=8.6471$ a.u.
The SSCHA is applied fully \abinitio{} with forces and
BO energies calculated using DFT within the
Perdew-Burke-Ernzerhof generalized gradient 
approximation~\cite{PhysRevLett.77.3865} 
and using ultrasoft
pseudopotentials as implemented in {\protect \sc
Quantum-ESPRESSO}~\cite {0953-8984-21-39-395502}.
A 60 Ry energy cutoff is used and a 
26 $\times $ 26 $\times $ 16 mesh for the
1BZ integrations in the unit cell.
Phonon frequencies and deformation potentials
are calculated
within linear response~\cite{0953-8984-21-39-395502,RevModPhys.73.515}
in a 6 $\times $ 6 $\times $ 4 $\mathbf{q}$ point grid.
For the SSCHA a 2 $\times $ 2 $\times $ 1 supercell 
containing 16 atoms is chosen. After applying symmetries, this gives us 
$N_p = 25$ parameters to be optimized. 
The difference between the SSCHA force-constants matrix
and the harmonic force-constants matrix in the 2 $\times $ 2 $\times $ 1
supercell is interpolated to the larger 6 $\times $ 6 $\times $ 4 supercell. 
Then, the harmonic 6 $\times $ 6 $\times $ 4 force-constants 
matrix is added to the result yielding the SSCHA force-constants
matrix in the larger 6 $\times $ 6 $\times $ 4 supercell.
The use of a smaller supercell for the SSCHA than for the harmonic case
is justified because the difference between the anharmonic and harmonic
force-constants matrices is localized in real space.  

The starting $\mathcal{H}_0$ Hamiltonian is the 
harmonic Hamiltonian and the starting number of
configurations $N_c = 20$. The initial
$\Phi(0)$ force-constants matrix of the 
2 $\times $ 2 $\times $ 1 supercell
does not have any imaginary eigenvalue. If it was the case,
the force-constants matrix should have been modified to avoid
initial imaginary frequencies.
The minimization is carried 
out with the following parameters: $\theta = 1$, 
$\zeta_c = 0.0003$ a.u.$^2$ and $\eta = 0.3$. 
The evolution of the SSCHA calculation is represented in
Fig. \ref{example-fig}. As it can be seen in Fig. \ref{example-fig}(a),
the calculation is stopped at $j=\mathrm{A}$ when 
$ | \langle \rho_{\mathcal{H}_j} / \rho_{\mathcal{H}_0} \rangle - 1 |
 \geq \eta$. Then, according to the flowchart in Fig. \ref{flowchart-fig},
a new set of configurations is created with $N_c = 20$ and the minimization
of the free energy continues. New sets of configurations are regenerated
at $j=\mathrm{B}$, $j=\mathrm{C}$ and $j=\mathrm{D}$. However, at step
$j=\mathrm{E}$ the calculation is stopped because all parameters 
in the gradient satisfy Eq. \eqref{conv-gradient} or Eq. \eqref{meaningfulness}
and $n_p=N_p$. The calculation could be stopped here, but
in order to increase the accuracy of the result
we generate another 380 configurations using 
$\mathcal{H}_{j={\mathrm{D}}}$ and calculate total energies
and atomic forces for them. Generating these configurations
with $\mathcal{H}_{j={\mathrm{D}}}$ instead of 
$\mathcal{H}_{j={\mathrm{E}}}$, we can recycle the previously generated
20 configurations. Thus, the number of configurations is increased
up to 400. Then, the minimization restarts till
$n_p=N_p$ at step $j=\mathrm{F}$. The 400 configurations were enough to
converge the phonon spectra.
Thus, in total 500 \abinitio{} force calculations
in supercells containing 16 atoms are needed. 
One should note that the bulk of the computational effort comes 
from the \abinitio{} calculation of the forces. The other operations 
of the SSCHA minimization require a negligible computational time.

In Fig. \ref{example-fig} we show as well
how the free energy is minimized
in the calculation. It is noteworthy that both
the harmonic contribution to the free energy, $F_{\mathcal{H}_j}$, and
the potential contribution, $\mathrm{tr} [ \rho_{\mathcal{H}_j}
( V - \mathcal{V}_j) ]$, vary a lot 
in the minimization. At the minimum,
at $j=\mathrm{F}$, the potential contribution to the 
free energy is not negligible with respect to the
harmonic contribution and it should be taken into account.
In Fig. \ref{example-fig}(b) the evolution of the phonon spectra
during the SSCHA minimization is illustrated. 

\subsection{Anharmonic phonon spectra of PtH at 100 GPa}
\label{phonon-pth}

\begin{figure}[t!]
\includegraphics[width=0.48\textwidth]{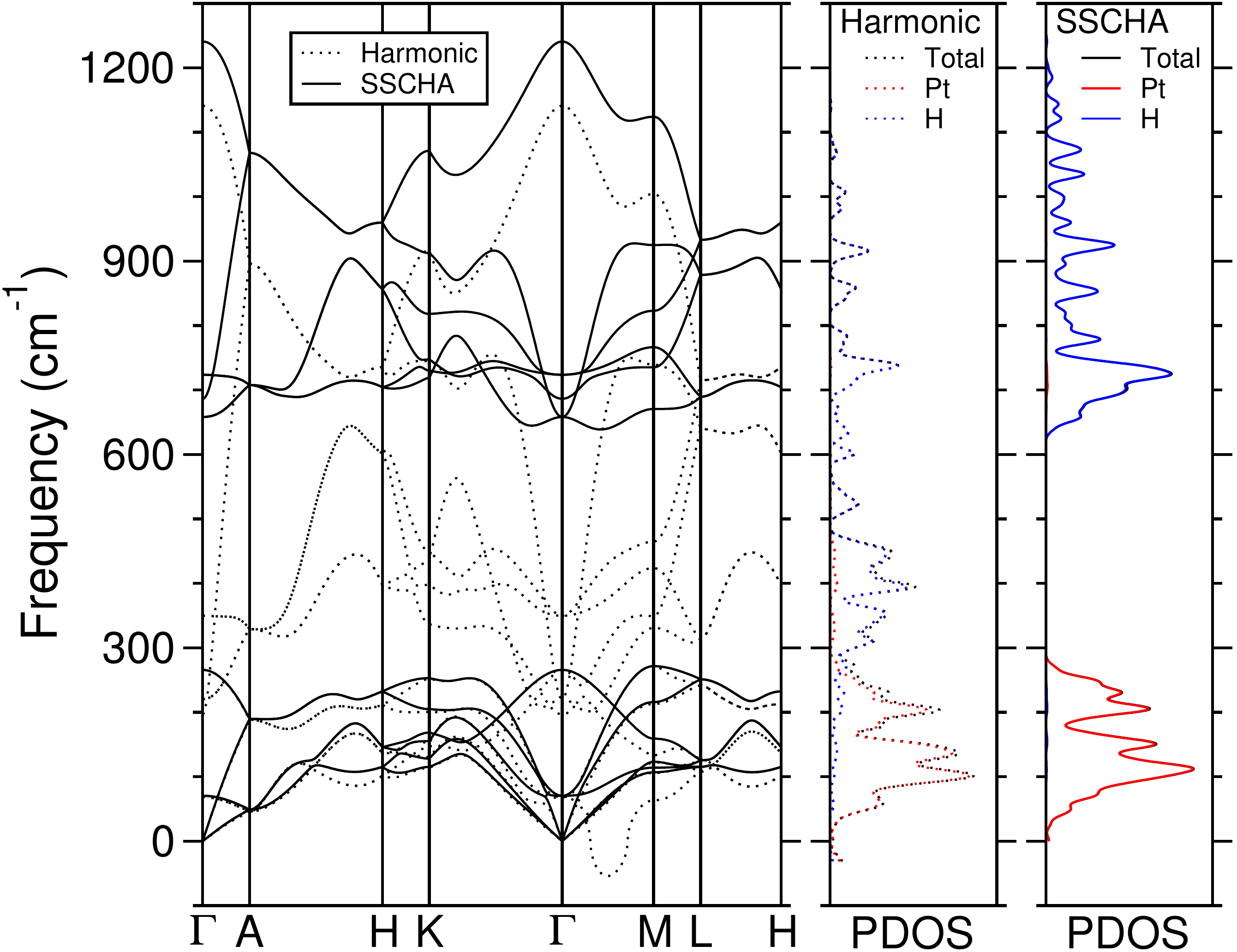}
\caption{ (Color online)
Phonon spectra and PDOS of PtH at 100 GPa and 0 K
in the harmonic approximation and in the SSCHA.
} 
\label{phonons-fig-pth}
\end{figure}

In Fig. \ref{phonons-fig-pth} we present the
phonon spectra calculated in the harmonic
approximation and in the SSCHA together
with the phonon density of states (PDOS).
The obtained harmonic phonon spectra 
is in good agreement with previous calculations
even if we observe a small instability along 
$\Gamma$M not present in previous 
calculations~\cite{PhysRevLett.107.117002,doi:10.1021/jp210780m}. 
The anharmonic correction of the phonon spectra
given by the SSCHA is huge. Even if all the
modes are affected by anharmonicity, the biggest
effect is attributed to the H-character modes 
with low energy in the harmonic approximation.
The small instability present in the harmonic
approximation completely disappears in the SSCHA. This 
suggests that the hcp phase of PtH is stabilized
by anharmonic effects down to the pressures where
it was observed in experiments~\cite{PhysRevB.83.214106}, 
even if harmonic calculations predict it to be unstable 
below 100 GPa~\cite{PhysRevLett.107.117002,doi:10.1021/jp210780m}.
This underlines that the SSCHA is a useful method
to treat systems that are apparently unstable in the
harmonic approximation but are stabilized by anharmonic
effects.    
  
It should be stressed that, even if in the harmonic approximation
a mixing between H and Pt character is observed in the
low-energy modes, the mixing is strongly suppressed
in the anharmonic case. This is evident in the 
projected PDOS shown in Fig. \ref{phonons-fig-pth}.
In order to be able to predict this reduction of the mixing,
the free energy must be minimized with respect to the
polarization vectors. 
It is unclear how methods like 
SCAILD~\cite{PhysRevLett.100.095901,Souvatzis2009888}
and the method presented by Antolin 
\etal{}~\cite{PhysRevB.86.054119}, which as far as we understand
do not optimize the polarization vectors,
behave in situations where the character of the
polarization vectors is strongly altered by anharmonicity
as in PtH. 

\subsection{Superconductivity of PtH at 100 GPa}
\label{elph-pth}

Once the phonon spectra renormalized by anharmonicity
has been obtained using the SSCHA, anharmonic effects
can be easily incorporated into the electron-phonon coupling
calculations. Assuming that the main effect of anharmonicity 
is a change in the phonon frequencies and polarizations, and that the 
deformation potential is unchanged, the anharmonic
Eliashberg function can be calculated as
\beqn
&& \alpha^{2}F(\omega) = \frac{1}{N(0) N_k N_q} \sum_{{\bf k}{\bf q}nm} 
  \sum_{\bar{s}\bar{t} \alpha \beta \mu} 
  \frac{\epsilon_{\mu \mathcal{H}}^{\bar{s} \alpha}(\mathbf{q}) 
           \epsilon_{\mu \mathcal{H}}^{\bar{t} \beta *}(\mathbf{q}) }{
        2 \omega_{\mu \mathcal{H}} (\mathbf{q}) \sqrt{M_{\bar{s}}M_{\bar{t}}}} 
   \times \nonumber \\
&&\ \  d^{\bar{s}\alpha}_{{\bf k}n,{\bf k}+{\bf q}m}
   d^{\bar{t}\beta *}_{{\bf k}n,{\bf k}+{\bf q}m}
   \delta(\epsilon_{{\bf k}n}) 
   \delta(\epsilon_{{\bf k+q}m})
   \delta(\omega -  \omega_{\mu \mathcal{H}} (\mathbf{q})).
\label{eliashberg}
\eeqn
In Eq. \eqref{eliashberg} $d^{\bar{s}\alpha}_{{\bf k}n,{\bf k}+{\bf q}m} = \bra{{\bf k}n}
\delta V / \delta u^{\bar{s}\alpha}(\mathbf{q}) \ket{{\bf k}+{\bf q}m}$
is the deformation potential, $\ket{{\bf k}n}$ is a  
Kohn-Sham state with energy $\epsilon_{{\bf k}n}$ 
measured from the Fermi  level ($\epsilon_F$), 
$N_k$ and $N_q$ are the number of electron
and phonon momentum points used for the 1BZ sampling, 
and $N(0)$ is the density of states per spin at $\epsilon_F$.
Note that in  Eq. \eqref{eliashberg} the sum over 
atomic indices is limited to the unit cell so that
phonon frequencies and polarization vectors are labeled
with a momentum $\mathbf{q}$. Similarly,
$ u^{\bar{s}\alpha}(\mathbf{q})$ is the Fourier transform
of  $u^{s\alpha}$.
We have used a finer 60 $\times$ 60 $\times$ 36 mesh
in the sum over $\mathbf{k}$ points in Eq. \eqref{eliashberg}. 

\begin{figure}[t!]
\includegraphics[width=0.48\textwidth]{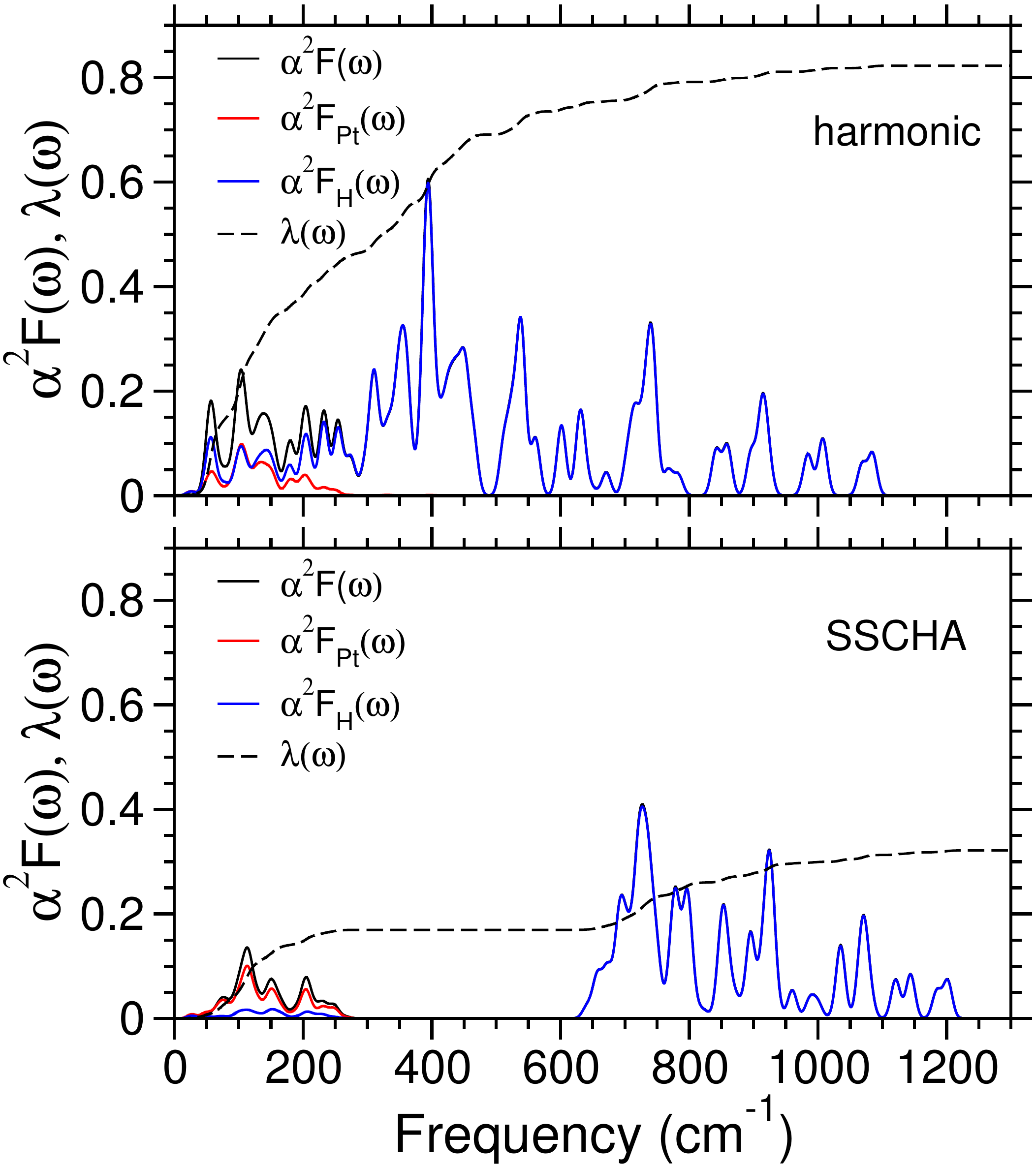}
\caption{ (Color online)
$\alpha^{2}F(\omega)$ and $\lambda(\omega)$ 
of PtH at 100 GPa and 0 K
in the harmonic approximation (top panel) and in the 
SSCHA (bottom panel). The contribution of Pt and H
atoms to $\alpha^{2}F(\omega)$ is depicted as well.
} 
\label{elph-fig-pth}
\end{figure}

\begin{table}[t]
\caption{Calculated $\lambda$, $\omega_{\mathrm{log}}$ and $T_c$ values for PtH
at 100 GPa and 0 K in the harmonic approximation and in the SSCHA.
}
\begin{center}
\begin{tabular*}{0.49\textwidth}{c c c c c } 
\hline
\hline
& $\lambda$ & $\omega_{\mathrm{log}}$(meV) &  $T_c [\mu^*=0.10]$(K) & $T_c [\mu^*=0.13]$(K) \\ 
\hline
Harmonic  & 0.82 & 25.3 & 14.5 & 11.8 \\
SSCHA     & 0.32 & 36.1 &  0.4 &  0.1 \\
\hline
\hline
\end{tabular*}
\end{center}
\label{table-tc}
\end{table}

The Eliashberg function in the harmonic approximation and in the SSCHA
is shown in Fig. \ref{elph-fig-pth} together with the
integrated electron-phonon coupling constant
\beq
\lambda(\omega) = 2 \int_0^{\omega} \mathrm{d}\omega' \frac{\alpha^2F(\omega')}{\omega'}.
\label{lambda-int}
\eeq
With the electron-phonon coupling constant $\lambda$, $\lambda = \lim_{\omega \to \infty} 
\lambda(\omega)$, and the logarithmic frequency average,
\beq
\omega_{\mathrm{log}} = \exp \left( \frac{2}{\lambda} 
\int_0^{\infty} \mathrm{d}\omega \frac{\alpha^2F(\omega)}{\omega}
\ln \omega \right),
\label{wlog}
\eeq 
we estimate $T_c$ making use of the Allen-Dynes modified McMillan
equation~\cite{PhysRevB.12.905},
using $\mu^*=0.10$ and $\mu^*=0.13$ for the Coulomb
pseudopotential. The results for $\lambda$, 
$\omega_{\mathrm{log}}$ and $T_c$ are summarized in Table
\ref{table-tc}.
In the harmonic approximation, despite the instability 
that barely contributes to 
$\alpha^{2}F(\omega)$, 
we obtain a $\lambda$ and $T_c$ in agreement with previous
calculations~\cite{PhysRevLett.107.117002}.
In the SSCHA, $\lambda$ is strongly suppressed due to the
enhancement of the frequencies induced by anharmonicity.
In Fig. \ref{elph-fig-pth} we show that the H contribution
to the Eliashberg function shifts to higher energies 
in the SSCHA, highly reducing the H contribution to $\lambda$.
While in the harmonic approximation H remarkably contributes 
to $\alpha^{2}F(\omega)$ at low energies, this is no longer
true in the anharmonic case. This again evidences the fact that
the H and Pt mixing of the low-energy modes disappears with
anharmonicity.

The suppression in $\lambda$ makes $T_c$ smaller than 1 K.
This means that the superconducting critical temperature
is reduced by an order of magnitude in PtH when anharmonicity
is included. Even if we do not include anharmonic corrections
in the deformation potential, our results indicate
that at 100 GPa PtH is not superconducting at around
17 K as measured in the experiment of 
silane~\cite{Eremets14032008}. The interpretation
that in the experiment in Ref.~\onlinecite{Eremets14032008} 
the superconductivity of PtH was measured is therefore
questioned by our calculations.

%%%%%%%%%%%%%%%%%%%%%%
% Application to PdH %
%%%%%%%%%%%%%%%%%%%%%%

\section{Application of the stochastic self-consistent harmonic 
approximation to palladium hydrides}
\label{pdh}

Secondly, we apply the
SSCHA to the strongly anharmonic palladium hydrides. In palladium hydrides
the anharmonic correction of the phonon frequencies   
is larger than the harmonic frequencies themselves, invalidating any 
perturbative approach as we have demonstrated recently
in Ref.~\onlinecite{PhysRevLett.111.177002}. The 
harmonic approximation displays
imaginary phonon frequencies for lattice parameters larger than 
approximately 7.72 a.u. Considering that experimental
lattice parameters of palladium hydrides are around 
7.73 a.u.~\cite{PhysRevB.12.117,PhysRevB.58.2591}, the quasiharmonic
approximation is not valid to study thermodynamic properties as the 
harmonic energy has no lower bound in case imaginary phonons are present.
Moreover, the harmonic approximation strongly overestimates superconducting
transition temperatures in palladium hydrides and, obviously, 
does not explain the inversion of the isotope 
effect~\cite{stritzker257,PhysRevB.10.3818,Schirber1984837}.
In Ref.~\onlinecite{PhysRevLett.111.177002} we showed how the SSCHA
explains the dynamical stability of palladium hydrides, 
the thermal expansion and even the inverse isotope effect.
Here we describe in further detail the thermodynamic properties of PdH,
PdD and PdT.  
  
%
%---------------------------
%
 
\subsection{Calculation details}
\label{details-pdh}

In the calculations presented here we make use of a model
potential built on top
of first-principles calculations that combines
the \abinitio{} harmonic potential with a fourth-order on-site
anharmonic potential. The reader is referred to 
Ref.~\onlinecite{PhysRevLett.111.177002} for the details
of the model potential and the SSCHA calculation.
The model potential allows us to reduce the statistical noise
in the calculation of the free energy.
We calculate the free energy of PdH, PdD and PdT
at several volumes and temperatures. 
The vibrational contribution to the free energy is a 
smooth function of the volume that can be fitted
accurately to a low order polynomial. We use a 
second order polynomial to fit these contributions. 
Then, the electronic ground state energy is added to 
the vibrational free energy to obtain the total
free energy. From the minimum of the
free energy at each temperature we calculate the 
dependence of the lattice parameter as a function of
temperature and
the value of the free energy at zero pressure
for each isotope.

%
%---------------------------
%
 
\subsection{Thermodynamic properties}
\label{thermodynamic}

\begin{figure}[t!]
\includegraphics[width=0.43\textwidth]{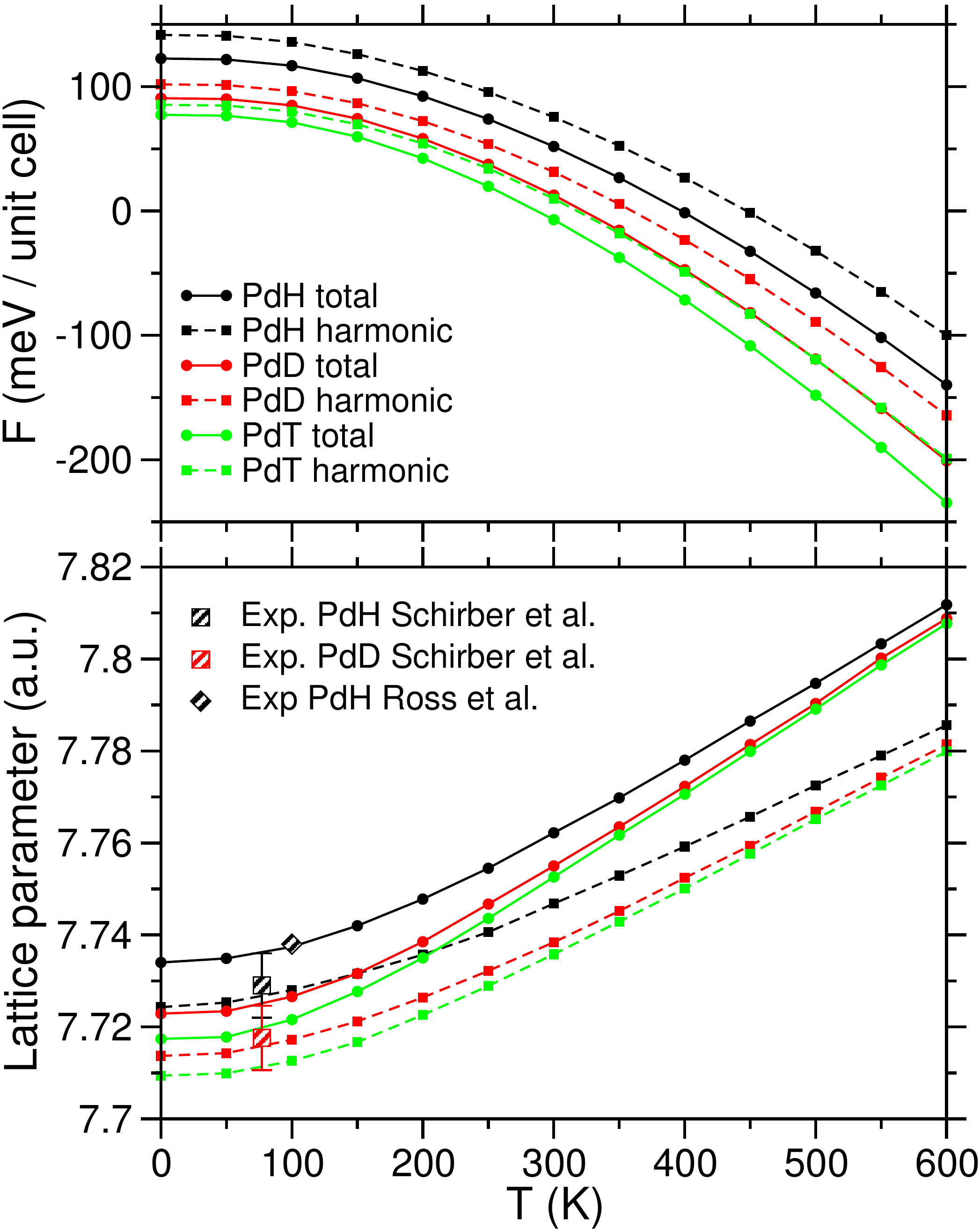}
\caption{ (Color online)
Total free energy at zero pressure (top panel) 
and the lattice parameter (bottom panel) 
as a function of temperature for
PdH, PdD and PdT. 
In both figures circles with solid lines represent
data obtained with $\mathcal{F}_H[\mathcal{H}]$ for the 
total vibrational contribution, 
while squares with dashed lines
represent data with exclusively the harmonic contribution
$F_{\mathcal{H}}$ for the vibrational contribution. 
Experimental lattice parameters 
obtained by Schirber \etal~\cite{PhysRevB.12.117} and
Ross \etal~\cite{PhysRevB.58.2591} are included.
} 
\label{free-energy-fig-pdh}
\end{figure}

In Fig. \ref{free-energy-fig-pdh} we plot the total free energy
at zero pressure and the equilibrium lattice parameter
for PdH, PdD and PdT as a function of temperature.
As it was shown in Ref.~\onlinecite{PhysRevLett.111.177002}, the results obtained
for the lattice parameter 
are in close agreement with experiments~\cite{PhysRevB.12.117,PhysRevB.58.2591}. 
Here we note that in the case of palladium hydrides
the inclusion of $\mathrm{tr} [ \rho_{\mathcal{H}}
( V - \mathcal{V}) ]$ in Eq. \eqref{var-free-energy2}, 
which gives the vibrational
free energy, is crucial to account for thermodynamic
properties. As it happens in PtH (see Fig. \ref{example-fig})
at the minimum this potential
contribution is not negligible. Indeed, as shown in Fig.
\ref{free-energy-fig-pdh} the total free energy 
is systematically overestimated if it is
calculated considering exclusively 
$F_{\mathcal{H}}$ for the vibrational contribution.
The overestimation is more important the higher the temperature
and the lighter the isotope.
In the lattice parameter the effect is very remarkable
since at 600 K, for example, neglecting 
the potential contribution $\mathrm{tr} [ \rho_{\mathcal{H}}
( V - \mathcal{V}) ]$ 
the lattice parameter is underestimated   
by 0.026 a.u. for PdH. Thus,
we can conclude that in palladium hydrides
neglecting $\mathrm{tr} [ \rho_{\mathcal{H}}
( V - \mathcal{V}) ]$ is not a 
good approximation. For strongly anharmonic
systems a similar behavior is expected. 

We should note that, as far as we understand, the SCAILD
method~\cite{PhysRevLett.100.095901,Souvatzis2009888}
and the method presented by Antolin \etal{}~\cite{PhysRevB.86.054119} 
do not include the $\mathrm{tr} [ \rho_{\mathcal{H}}
( V - \mathcal{V}) ]$ potential contribution in the free energy.
On the contrary, in the SSCHA we can accurately
calculate this contribution without
applying thermodynamic integration, which is
the way it can be incorporated  
in the TDEP method~\cite{PhysRevB.87.104111}. 

%%%%%%%%%%%%%%%
% CONCLUSIONS %
%%%%%%%%%%%%%%%

\section{Conclusions}
\label{conclusions}

The development of a non-perturbative treatment of anharmonic corrections 
to the phonon frequencies and the free energy is a major challenge with impacts in 
many domains of physics and chemistry, including superconductivity,
charge-density waves, thermoelectric materials, ferroelectrics, 
thermodynamic phase transitions and many more.
In this work we solve this issue by developing the stochastic 
self-consistent harmonic approximation. This method
is non-perturbative and scales favorably with the system size
as it can benefit from the locality of the anharmonic contribution to the forces
acting on the atoms. 
It is variational in the free energy at any
temperature as it minimizes it with respect to
a trial harmonic density matrix. 
The gradient of the free energy is calculated as a function of all
the independent parameters in the trial harmonic
Hamiltonian, and, thus, the SSCHA can
calculate the equilibrium positions, phonon frequencies
and polarizations vectors beyond
perturbation theory. The  
gradient of the free energy is calculated stochastically from
total energies and atomic forces on supercells
with ionic configurations described by the probability
distribution defined by the trial Hamiltonian.
Therefore, the SSCHA can be applied at different
degrees of theory using different total-energy-force
engines, e.g., classical potentials, \abinitio{} DFT approaches, 
or Quantum Monte Carlo (QMC). 
The SSCHA scheme is devised to minimize the number of calls
to the  total-energy-force engine.
The algorithm is such that all the computer time goes
in the calculation of total energies and forces if
a DFT approach or a more precise approach like QMC
is followed.

We apply the method to platinum hydride
at high-pressure and palladium hydrides. In the case
of palladium hydrides we demonstrate how within the SSCHA
we can calculate thermodynamic properties in agreement with 
experiments even when the 
quasiharmonic approximation breaks down. 
In PtH, we have reanalyzed the phonon spectra and its superconducting
properties at 100 GPa. We have shown that anharmonicicty 
strongly renormalizes the phonon frequencies beyond the perturbative
regime with a considerable suppression of the superconducting critical
temperature with respect to previous harmonic 
estimates~\cite{PhysRevLett.107.117002,PhysRevB.83.214106,PhysRevB.84.054543}. 
This result makes us wonder whether the interpretation that in the
experiment in silane by Eremets \etal{}~\cite{Eremets14032008} 
the superconductivity of
PtH was measured~\cite{PhysRevB.83.214106,PhysRevB.84.054543,PhysRevLett.107.117002} 
is valid. Considering the similar suppression in palladium hydrides,
which explains its inverse isotope effect~\cite{PhysRevLett.111.177002}, and 
the lack of superconductivity measured in AlH$_3$ despite been
predicted within the harmonic 
theory~\cite{PhysRevLett.100.045504,PhysRevB.82.104504}, 
it seems that anharmonicity might strongly affect the
predicted $T_c$ values in several  
hydrides~\cite{Kim16022010,
PhysRevB.82.104504,PhysRevLett.107.117002,Gao26012010,gao:107002,
cudazzo:257001,martinez-canales:087005}.

%
% Acknowledgements   
%

\section*{ACKNOWLEDGMENTS}

I.E. would like to acknowledge financial support
from the 
Department of Education, Language Policy and Culture of the Basque
Government (Grant No. BFI-2011-65). 
The authors acknowledge support from the Graphene Flagship and
from the French state funds managed by
the ANR within the Investissements d'Avenir programme under reference
ANR-11-IDEX-0004-02, ANR-11-BS04-0019 and ANR-13-IS10-0003-01.
Computer facilities were provided by CINES, CCRT and IDRIS
(Project No. x2014091202).

%%%%%%%%%%%%%%
% APPENDICES %
%%%%%%%%%%%%%

\appendix

%
%-----------------------
%

\section{Calculation of the gradient of the free energy}
\label{app-gradient}

In order to calculate the derivatives of 
$\mathcal{F}_H[\mathcal{H}]$, it is convenient to work with normal
coordinates. The quantum statistical average of an operator that exclusively
depends on atomic positions can be calculated in normal coordinates
applying the change of variables defined in Eq. \eqref{qmu} to Eq.
\eqref{quantum-average}, namely
\beq
\mathrm{tr} [\rho_{\mathcal{H}} O] 
    = \int \mathrm{d}\mathbf{q} O(\dots, R_{{\rm eq}}^{s \alpha} + 
                                         \sum_{\mu=1}^{3N} \frac{1}{\sqrt{M_s}} 
                                         \epsilon_{\mu \mathcal{H}}^{s \alpha} q_{\mu},
                                   \dots)
         \rho_{\mathcal{H}}(\mathbf{q}),      
\label{quantum-average-q}
\eeq
where $\mathbf{q}$ is a general configuration of the normal coordinates  
and $\rho_{\mathcal{H}}(\mathbf{q})$ is the probability to find the system in the 
configuration $\mathbf{q}$. In normal coordinates, 
$\rho_{\mathcal{H}}(\mathbf{q})$ is nothing but a product of 
Gaussians~\cite{doi:10.1021/jp044251w}:
\beq
\rho_{\mathcal{H}}(\mathbf{q})=
                   \prod_{\mu=1}^{3N} \frac{1}{a_{\mu \mathcal{H}}\sqrt{2 \pi} }
                    e^{ - \frac{q^2_{\mu}}{2 a^2_{\mu \mathcal{H}}}}. 
\label{probability-q}
\eeq 
If we make use of the $y_{\mu} = q_{\mu} / a_{\mu \mathcal{H}}$ 
change of variables, the integral can be written as
\beqn
\mathrm{tr} [\rho_{\mathcal{H}} O] 
    & = & \int \mathrm{d}\mathbf{y} O(\dots, R_{{\rm eq}}^{s \alpha} + 
                                   \sum_{\mu=1}^{3N} \frac{1}{\sqrt{M_s}} 
                                   \epsilon_{\mu \mathcal{H}}^{s \alpha} 
                                    a_{\mu \mathcal{H}} y_{\mu},
                                   \dots)  \nonumber \\ && \times
                            \prod_{\mu=1}^{3N} \frac{1}{\sqrt{2 \pi} }
                    e^{ - \frac{y^2_{\mu}}{2 }},      
\label{quantum-average-y}
\eeqn
where $\mathbf{y}$ represents all the set of $y_{\mu}$'s.
Writing the quantum statistical average in this way,
the exponential part becomes independent of phonon frequencies, polarizations
vectors and equilibrium positions.

The only non analytic part in the calculation of 
$\mathcal{F}_H[\mathcal{H}] = F_{\mathcal{H}} + \mathrm{tr} [ \rho_{\mathcal{H}}
      ( V - \mathcal{V}) ]$
is the quantum statistical average of the potential, the
$\mathrm{tr} [ \rho_{\mathcal{H}} V ]$ term. The analytic expression of 
$F_{\mathcal{H}}$ is given in Eq. \eqref{f0} and
\beq
\mathrm{tr} [ \rho_{\mathcal{H}} \mathcal{V} ] = \sum_{\mu=1}^{3N} 
    \frac{1}{2} \omega^2_{\mu \mathcal{H}} a^2_{\mu \mathcal{H}} .
\label{av-harmonic-potential}
\eeq
Before we demonstrate Eqs. \eqref{gradient-eq} and \eqref{gradient-phi},
we also would like to note the following analytical relation involving
the $\mathbf{f}_{\mathcal{H}}(\mathbf{R})$ harmonic forces:
\beqn
&& \int \mathrm{d}\mathbf{R} f^{s \alpha}_{\mathcal{H}}(\mathbf{R})
     (R^{t\beta} - R^{t\beta}_{\mathrm{eq}}) \rho_{\mathcal{H}}(\mathbf{R}) = 
\nonumber \\ && \ \ \     -  \sum_{\mu=1}^{3N} \sqrt{\frac{M_s}{M_t}} 
        \epsilon_{\mu \mathcal{H}}^{s \alpha}
        \epsilon_{\mu \mathcal{H}}^{t \beta}
        \omega^2_{\mu \mathcal{H}} a^2_{\mu \mathcal{H}}.
\label{analytic-harmonic-force}
\eeqn

First of all, if we write $\mathrm{tr} [\rho_{\mathcal{H}} V]$ as in Eq.
\eqref{quantum-average-y}, it is easy to observe that
\beq
\boldsymbol{\nabla}_{\mathbf{R_{\mathrm{eq}}}}\mathcal{F}_H[\mathcal{H}] =
\boldsymbol{\nabla}_{\mathbf{R_{\mathrm{eq}}}} \mathrm{tr} [\rho_{\mathcal{H}} V] =
  -  \int \mathrm{d}\mathbf{R} \mathbf{f}(\mathbf{R}) 
                     \rho_{\mathcal{H}}(\mathbf{R}).
\label{gredient-eq-app}
\eeq
Noting that $\int \mathrm{d}\mathbf{R} \mathbf{f}_{\mathcal{H}}(\mathbf{R}) 
                     \rho_{\mathcal{H}}(\mathbf{R}) = 0$,
we recover Eq. \eqref{gradient-eq}. 

Secondly, we calculate $\boldsymbol{\nabla}_{\Phi} \mathcal{F}_H[\mathcal{H}]$ using the
chain rule as
\beq
\boldsymbol{\nabla}_{\Phi} \mathcal{F}_H[\mathcal{H}] =
   \sum_{\mu} \frac{\partial \mathcal{F}_H[\mathcal{H}]}{\partial a_{\mu \mathcal{H}}}
   \boldsymbol{\nabla}_{\Phi} a_{\mu \mathcal{H}} 
 +  \sum_{\mu s \alpha}  \frac{\partial \mathcal{F}_H[\mathcal{H}]}{\partial 
                              \epsilon^{s \alpha}_{\mu \mathcal{H}}}
   \boldsymbol{\nabla}_{\Phi}  \epsilon^{s \alpha}_{\mu \mathcal{H}}. 
\label{chain-rule}
\eeq
The partial derivative with respect to the polarization vectors
is easily obtained writing again
$\mathrm{tr} [\rho_{\mathcal{H}} V]$ as in Eq. \eqref{quantum-average-y}.
In particular,
\beqn
\frac{\partial \mathcal{F}_H[\mathcal{H}]}{\partial 
                              \epsilon^{s \alpha}_{\mu \mathcal{H}}} & = & 
\frac{\partial \mathrm{tr} [\rho_{\mathcal{H}} V] 
}{\partial \epsilon^{s \alpha}_{\mu \mathcal{H}}} = 
- \sum_{t \beta} \sqrt{\frac{M_t}{M_s}} \epsilon_{\mu \mathcal{H}}^{t \beta} \nonumber \\ 
& \times & \int \mathrm{d}\mathbf{R} f^{s \alpha}(\mathbf{R})
     (R^{t\beta} - R^{t\beta}_{\mathrm{eq}}) \rho_{\mathcal{H}}(\mathbf{R}).
\label{partial-pol-vec}
\eeqn 
Making use of Eq. \eqref{analytic-harmonic-force} we can write
\beqn
&& \frac{\partial \mathcal{F}_H[\mathcal{H}]}{\partial 
                              \epsilon^{s \alpha}_{\mu \mathcal{H}}} = 
    \epsilon^{s \alpha}_{\mu \mathcal{H}} \omega^2_{\mu \mathcal{H}} a^2_{\mu \mathcal{H}} 
- \sum_{t \beta} \sqrt{\frac{M_t}{M_s}} \epsilon_{\mu \mathcal{H}}^{t \beta} \nonumber \\ 
&& \ \  \times  \int \mathrm{d}\mathbf{R} 
     [ f^{s \alpha}(\mathbf{R}) - f^{s \alpha}_{\mathcal{H}}(\mathbf{R})]
     (R^{t\beta} - R^{t\beta}_{\mathrm{eq}}) \rho_{\mathcal{H}}(\mathbf{R}).
\label{partial-pol-vec-2}
\eeqn 
On the other hand,
%The partial derivative with respect to the normal length can be calculated similarly
%as
\beq
\frac{\partial \mathcal{F}_H[\mathcal{H}]}{\partial a_{\mu \mathcal{H}}}
=  \frac{\partial \left( F_{\mathcal{H}} - \mathrm{tr} 
    [ \rho_{\mathcal{H}} \mathcal{V} ] \right)
   }{\partial a_{\mu \mathcal{H}}} + 
   \frac{\partial \mathrm{tr} [ \rho_{\mathcal{H}} V]}{\partial a_{\mu \mathcal{H}}}
\label{partial-a-1}
\eeq
It is straightforward to demonstrate that
\beqn
\frac{\partial \mathrm{tr} [ \rho_{\mathcal{H}} V]}{\partial a_{\mu \mathcal{H}}}
& = & -  \sum_{ s t \alpha \beta} \sqrt{\frac{M_t}{M_s}}
        \epsilon_{\mu \mathcal{H}}^{s \alpha}
        \epsilon_{\mu \mathcal{H}}^{t \beta} \frac{1}{a_{\mu \mathcal{H}}}
      \nonumber \\ & \times &
      \int \mathrm{d}\mathbf{R} f^{s \alpha}(\mathbf{R})
     (R^{t\beta} - R^{t\beta}_{\mathrm{eq}}) \rho_{\mathcal{H}}(\mathbf{R})
\label{partial-a-2}
\eeqn
if Eq. \eqref{quantum-average-y} is used to write the quantum statistical
average. Using Eq. \eqref{analytic-harmonic-force} the equation above
can be rewritten as
\beqn
&& \frac{\partial \mathrm{tr} [ \rho_{\mathcal{H}} V]}{\partial a_{\mu \mathcal{H}}}
 = \omega^2_{\mu \mathcal{H}} a_{\mu \mathcal{H}} -  \sum_{ s t \alpha \beta} 
        \sqrt{\frac{M_t}{M_s}}
        \epsilon_{\mu \mathcal{H}}^{s \alpha}
        \epsilon_{\mu \mathcal{H}}^{t \beta} \frac{1}{a_{\mu \mathcal{H}}}
      \nonumber \\ &&  \times 
     \int \mathrm{d}\mathbf{R} 
       [ f^{s \alpha}(\mathbf{R}) - f^{s \alpha}_{\mathcal{H}}(\mathbf{R})]
     (R^{t\beta} - R^{t\beta}_{\mathrm{eq}}) \rho_{\mathcal{H}}(\mathbf{R}).
\label{partial-a-3}
\eeqn
Finally, plugging Eqs. \eqref{partial-a-3} and \eqref{partial-pol-vec-2}
into Eq. \eqref{chain-rule}, and noting that 
\beqn
\frac{\partial \left( F_{\mathcal{H}} - \mathrm{tr} [ \rho_{\mathcal{H}} \mathcal{V} ] \right)
   }{\partial a_{\mu \mathcal{H}}} + \omega^2_{\mu \mathcal{H}} a_{\mu \mathcal{H}} & = & 0 
   \label{cancellation1}\\
\sum_{s \alpha} \epsilon^{s \alpha}_{\mu \mathcal{H}}
\boldsymbol{\nabla}_{\Phi}  \epsilon^{s \alpha}_{\mu \mathcal{H}} & = & 0,
\label{cancellation2}
\eeqn
it is straightforward to obtain the expression for
$\boldsymbol{\nabla}_{\Phi} \mathcal{F}_H[\mathcal{H}]$ 
given in Eq. \eqref{gradient-phi}.

The only elements contributing to $\boldsymbol{\nabla}_{\Phi} \mathcal{F}_H[\mathcal{H}]$ 
that have not been specified yet are $\boldsymbol{\nabla}_{\Phi} a_{\mu \mathcal{H}}$ and
$\boldsymbol{\nabla}_{\Phi}  \epsilon^{s \alpha}_{\mu \mathcal{H}}$.
Considering how eigenvalues and eigenvectors of a matrix are modified
when the matrix itself is varied, we can get the following 
expressions for the components of these gradients:
\beqn
\frac{\partial a_{\mu \mathcal{H}}}{\partial \Phi^{\alpha \beta}_{st}} & = &
  \frac{\partial a_{\mu \mathcal{H}}}{\partial \omega_{\mu \mathcal{H}}}
  \frac{1}{2 \omega_{\mu \mathcal{H}}} \frac{ \epsilon_{\mu \mathcal{H}}^{s \alpha}
        \epsilon_{\mu \mathcal{H}}^{t \beta}}{\sqrt{M_sM_t}} 
\label{dadphi} \\
\frac{\partial \epsilon^{s \alpha}_{\mu \mathcal{H}}}{\partial 
   \Phi^{\gamma \beta}_{kt}} & = &
  \sum_{\nu, \nu \neq \mu} \frac{ \epsilon_{\nu \mathcal{H}}^{k \gamma}
                                  \epsilon_{\mu \mathcal{H}}^{t \beta} }{
                 \sqrt{M_kM_t} ( \omega^2_{\mu \mathcal{H}} - \omega^2_{\nu \mathcal{H}}) }
       \epsilon_{\nu \mathcal{H}}^{s \alpha}. 
\label{dedphi}
\eeqn
The $\partial a_{\mu \mathcal{H}} / \partial \omega_{\mu \mathcal{H}}$ derivative
in Eq. \eqref{dadphi} can be obtained from Eq. \eqref{normal-length}.

%
%-----------------------
%

\section{Symmetrization and reduction of the basis for the
force-constants matrix}
\label{app-symmetry}

In order to apply $\hat{S}$ and time-reversal symmetries
to the $\{\mathcal{G}_{\mathrm{(ns)}}(m) \}_{m=1,\dots,(3n)^2 N_1 N_2 N_3}$ 
basis, it is convenient to to work in the 
unit-cell phonon-momentum $\mathbf{q}$ space.
This is so because any $\Phi$ force-constants matrix of this vector space
respects translational symmetries and, thus,
its Fourier transform can be labeled with 
a single $\mathbf{q}$ vector of the 1BZ of the unit cell.
The number of $\mathbf{q}$ points in the 1BZ 
is determined by the supercell size. A $ N_1 \times N_2 \times N_3$ supercell
means a $N_1 \times N_2 \times N_3$ phonon-momentum grid in the 1BZ
of the unit cell. The Fourier
components of the real-space force-constants matrix of the supercell
$\Phi$ are labeled as $\Phi(\mathbf{q})$. Each $\Phi(\mathbf{q})$
is a $3n \times 3n$ Hermitian matrix. 
Therefore, in practice we do not work with the vector space 
defined by the
$\{\mathcal{G}_{\mathrm{(ns)}}(m) \}_{m=1,\dots,(3n)^2 N_1 N_2 N_3}$
basis, but with the vector space formed by the $3n \times 3n$
Hermitian matrices. The dimension of this vector space is
$(3n)^2$ and let $\{ \bar{\mathcal{G}}_{\mathrm{(ns)}}(\sigma) 
\}_{\sigma=1,\dots,(3n)^2}$ be an orthonormal basis of it. 
The elements of the basis satisfy the orthonormality
condition defined in Eq. \eqref{orthonormality-mat}.
Therefore, any $\Phi(\mathbf{q})$ can be decomposed in this basis as
\beq
\Phi(\mathbf{q}) = \sum_{\sigma = 1}^{(3n)^2} c_{\mathrm{(ns)}} (\mathbf{q},\sigma)
                   \bar{\mathcal{G}}_{\mathrm{(ns)}}(\sigma),
\label{decompose-mat-ns}
\eeq
where the $c_{\mathrm{(ns)}} (\mathbf{q},\sigma)$ coefficients determine
the value of $\Phi(\mathbf{q})$.
 
Any $\Phi(\mathbf{q})$ matrix described in the 
 $\{ \bar{\mathcal{G}}_{\mathrm{(ns)}}(\sigma) 
\}_{\sigma=1,\dots,(3n)^2}$ basis as in Eq. \eqref{decompose-mat-ns}
transforms under the symmetry operations $\hat{S}$
of the space group as
\beq
\Phi(S\mathbf{q}) = T_{\hat{S}}(\mathbf{q}) \Phi(\mathbf{q}) 
                    T_{\hat{S}}^{\dagger}(\mathbf{q}), 
\label{dynmatsym}
\eeq
where the unitary $T_{\hat{S}}(\mathbf{q})$
matrices are given in Refs.~\onlinecite{RevModPhys.40.1,RevModPhys.40.38,
Hendrikse1995297}. Eq. \eqref{dynmatsym}
shows that many of the 
$\mathbf{q}$ points in the  $N_1 \times N_2 \times N_3$ mesh
are equivalent by symmetry since the force-constants
matrices at $S\mathbf{q}$ points of the 1BZ are related by symmetry
to the force-constants matrix at $\mathbf{q}$.
The set of symmetry related $S\mathbf{q}$ points is named as the star of 
$\mathbf{q}$ and we denote it as $\{\mathbf{q}^*\}$.
The $\mathbf{q}$ points not related by Eq.
\eqref{dynmatsym} form the irreducible
Brillouin zone (IBZ). Therefore, we can
restrict
the $\Phi(\mathbf{q})$ Fourier-components of the supercell force-constants matrix at 
the $\mathbf{q}$ points in the IBZ. Indeed, all the 
$N_1 \times N_2 \times N_3$  $\Phi(\mathbf{q})$ matrices can be
generated by symmetry using Eq. \eqref{dynmatsym}.

Eq. \eqref{dynmatsym} can be used to symmetrize the 
elements of the $\{ \bar{\mathcal{G}}_{\mathrm{(ns)}}(\sigma) 
\}_{\sigma=1,\dots,(3n)^2}$ basis with respect to
the $\hat{S}$ operations and time-reversal. For instance,
if 
\beq
S_{\mathbf{q}}     \mathbf{q} = \mathbf{q} + \mathbf{G}_{\mathbf{q}} 
\label{symq} 
\eeq
or
\beq
S_{-\mathbf{q}}     \mathbf{q} = - \mathbf{q} + \mathbf{G}_{-\mathbf{q}},
\label{symminusq} 
\eeq
we can use the $\hat{S}_{\mathbf{q}}$ and $\hat{S}_{-\mathbf{q}}$
symmetry operations to symmetrize the basis. The $\hat{S}_{\mathbf{q}}$ symmetry
operations form the so-called small group of $\mathbf{q}$.
The $\hat{S}_{-\mathbf{q}}$ operations can be used to symmetrize
the basis because of time-reversal symmetry.
From Eq. \eqref{dynmatsym} it is straightforward
to observe that the
elements of the basis can be symmetrized as
\beqn
&& \bar{\mathcal{G}}_{\mathrm{(s)}}(\mathbf{q},\sigma) = \frac{1}{N_{S_{\mathbf{q}}}}
      \sum_{\hat{S}_{\mathbf{q}}} 
      T_{\hat{S}_{\mathbf{q}}}^{\dagger}(\mathbf{q}) 
      \bar{\mathcal{G}}_{\mathrm{(ns)}}(\sigma) 
      T_{\hat{S}_{\mathbf{q}}}(\mathbf{q})
      \nonumber \\ 
  && \ \ \ \ + \frac{1}{N_{S_{-\mathbf{q}}}}
      \sum_{\hat{S}_{-\mathbf{q}}} 
      T_{\hat{S}_{-\mathbf{q}}}^{\dagger}(\mathbf{q}) 
      \bar{\mathcal{G}}^*_{\mathrm{(ns)}}(\sigma) 
      T_{\hat{S}_{-\mathbf{q}}}(\mathbf{q}),
\label{sym-basis}
\eeqn
where $N_{S_{\mathbf{q}}}$ is the number symmetry operations
in the small group of $\mathbf{q}$ and
 $N_{S_{-\mathbf{q}}}$ the number of symmetry operations 
satisfying Eq. \eqref{symminusq}. 
We perform this symmetrization at each $\mathbf{q} \in$ IBZ. 
The $\{ \bar{\mathcal{G}}_{\mathrm{(s)}}(\mathbf{q},\sigma) \}_{\sigma=1,\dots,(3n)^2}$ 
basis becomes overcomplete after the symmetrization. 
We reduce the basis applying a Gram-Schmidt orthonormalization
procedure. We label the set of matrices that form the basis
of the symmetrized subspace as 
$\{ \bar{\mathcal{G}}(\mathbf{q},\sigma) \}_{\sigma=1,\dots,N_r(\mathbf{q})}$,
$N_r(\mathbf{q})$ being the dimension of the subspace for
$\mathbf{q} \in$ IBZ.

In this framework, it is also easy to impose the acoustic
sum rule (ASR) as among the $\bar{\mathcal{G}}_{\mathrm{(s)}}(\Gamma,\sigma)$ matrices
there must exist three translation generators, one for each
Cartesian direction: $\mathcal{G}_t^{x}$, $\mathcal{G}_t^{y}$ and 
$\mathcal{G}_t^{z}$. According to the ASR, the translation vectors
must be eigenvectors of the force-constants matrix at $\Gamma$
with vanishing eigenvalue. In order to impose this property, we redefine
all the generators at $\Gamma$ according to the
\beq
\bar{\mathcal{G}}_{\mathrm{ASR}}(\Gamma,\sigma) = (\mathbb{I}-\mathcal{G}_t)
              \bar{\mathcal{G}}_{\mathrm{(s)}}(\Gamma,\sigma)(\mathbb{I}-\mathcal{G}_t)
\label{impose-asr}
\eeq
projection, where $\mathbb{I}$ is the identity matrix and 
$\mathcal{G}_t=\mathcal{G}_t^{x}+\mathcal{G}_t^{y}+\mathcal{G}_t^{z}$.
Then, instead of performing the Gram-Schmidt orthonormalization
in the  $\{\bar{\mathcal{G}}_{\mathrm{(s)}}(\Gamma,\sigma)\}_{\sigma=1,\dots,(3n)^2}$
basis, we perform it in the   
$\{\bar{\mathcal{G}}_{\mathrm{ASR}}(\Gamma,\sigma)\}_{\sigma=1,\dots,(3n)^2}$
one. This procedure gives us the 
$\{ \bar{\mathcal{G}}(\Gamma,\sigma) \}_{\sigma=1,\dots,N_r(\mathbf{\Gamma})}$
basis of the symmetric subspace at $\Gamma$.

Once the symmetry reduced basis is found for all
$\mathbf{q} \in$ IBZ, we decompose the initial
force-constants matrix $\Phi(0)$ in the basis. In order to decompose
it, it is sufficient to decompose the Fourier transformed
force-constants matrices at $\mathbf{q} \in$ IBZ. 
The decomposition
of each of these matrices is given as
\beq
\Phi(\mathbf{q},0) = \sum_{\sigma=1}^{N_r(\mathbf{q})} 
          c_0(\mathbf{q},\sigma) \bar{\mathcal{G}}(\mathbf{q},\sigma)
          ,
\label{decompose-dyn}
\eeq
where the $c_0(\mathbf{q},\sigma)$
coefficients can be obtain from the 
\beq
c_0(\mathbf{q},\sigma) = \langle \Phi (\mathbf{q},0), 
\bar{\mathcal{G}}(\mathbf{q},\sigma) \rangle 
\label{get-coeff}
\eeq
scalar product.
The scalar product in Eq. \eqref{get-coeff} is defined in
Eq. \eqref{scalar-prod-mat} though, in this case, the sum
over atom indices is limited to the unit cell. 
It is clear that the total number of coefficients
needed to determine $\Phi(0)$, or in general any $\Phi(j)$,
is $N_R = \sum_{\mathbf{q} \in \mathrm{IBZ}} N_r(\mathbf{q})$.

Finally, in order to obtain Eq. \eqref{decompose-fc-stepj},
we need to Fourier transform to real space the 
$\bar{\mathcal{G}}(\mathbf{q},\sigma)$ matrices.
The $\bar{\mathcal{G}}(\mathbf{q}',\sigma)$
matrices for all $\mathbf{q}' \in \{\mathbf{q}^*\}$ 
can be obtained using Eq. 
\eqref{dynmatsym}. The 
\beq
\mathcal{G}_{st}^{\alpha \beta} (\mathbf{q},\sigma) = 
      \frac{1}{N_{q}}  \sum_{\mathbf{q}' \in \{\mathbf{q}^*\}} 
                      \bar{\mathcal{G}}_{\bar{s}\bar{t}}^{\alpha \beta}(\mathbf{q}',\sigma) 
                      e^{i \mathbf{q}' (\mathbf{R}_{s\bar{s}} - \mathbf{R}_{t\bar{t}})}
\label{fourier-back}
\eeq 
Fourier transform
gives us the value of the matrix in real space.
In Eq. \eqref{fourier-back} $N_{q}=N_1 \times N_2 \times N_3$
is the total number of $\mathbf{q}$ points in the 1BZ and
$\mathbf{R}_{s\bar{s}}$ is the lattice vector that connects
the $s$-th atom of the supercell with the equivalent atom $\bar{s}$
in the unit cell, $\mathbf{R}_{s\bar{s}} = \mathbf{R}^s_{{\rm eq}} - 
\mathbf{R}^{\bar{s}}_{{\rm eq}}$.
Note that $\mathcal{G}_{st}^{\alpha \beta} (\mathbf{q},\sigma)$
denotes an element of the symmetrized basis in real space and
that $\mathbf{q},\sigma$ is nothing but the label of a matrix
in the basis. Moreover, as for 
$\mathbf{q}' \in \{\mathbf{q}^*\}$ $c_0(\mathbf{q}',\sigma)
= c_0(\mathbf{q},\sigma)$, the $\Phi(0)$ force-constants
matrix in real space can be calculated as
\beq
\Phi_{st}^{\alpha \beta} (0) = \sum_{\mathbf{q} \in \mathrm{IBZ}} 
           \sum_{\sigma=1}^{N_r(\mathbf{q})} c_0(\mathbf{q},\sigma) 
           \mathcal{G}_{st}^{\alpha \beta} (\mathbf{q},\sigma).
\label{decompose-fc-app}
\eeq 
Simplifying the notation as $m \equiv \mathbf{q},\sigma$
we get
\beq
\Phi(0) = \sum_{m=1}^{N_R} c_0(m) 
          \mathcal{G}(m),
\label{decompose-fc-app-2}
\eeq
where the sum runs over the $N_R$ different coefficients.
Obviously at CG step $j$ Eq. \eqref{decompose-fc-app-2}
holds as in Eq. \eqref{decompose-fc-stepj}.  

%
%-----------------------
%

\section{Creating the set of ionic configurations}
\label{app-stat}

In order to create the $\{ \mathbf{R}_I \}_{I=1,\dots,N_c}$
set of ionic configurations according
to the probability distribution
$\rho_{\mathcal{H}}(\mathbf{R})$, we take advantage of the
Gaussian character of $\rho_{\mathcal{H}}(\mathbf{q})$
(see Eq. \eqref{probability-q}).
First of all,  
a set of $\{ y_{\mu I} \}_{I=1,\dots,N_c}$ 
configurations is created. Each $y_{\mu I}$ 
is a random number created according to a purely Gaussian distribution. 
Then, we multiply each $y_{\mu I}$ by
the corresponding $a_{\mu \mathcal{H}}$ normal length. 
This operation gives us a set of 
configurations for the normal coordinates described by the 
$\rho_{\mathcal{H}}(\mathbf{q})$
probability distribution: 
$\{ q_{\mu I} = a_{\mu \mathcal{H}} y_{\mu I} \}_{I=1,\dots,N_c}$. 
From Eq. \eqref{qmu}, these normal coordinates define
a set of configurations for the ionic positions 
$\{ \mathbf{R}_I \}_{I=1,\dots,N_c}$, where 
each component is given as
\beq
 R^{s \alpha}_I = R_{{\rm eq}}^{s \alpha} + 
                         \sum_{\mu=1}^{3N} \frac{1}{\sqrt{M_s}} 
                         \epsilon_{\mu \mathcal{H}}^{s \alpha} 
                         a_{\mu \mathcal{H}} y_{\mu I}  
                         .
\label{ionic-configurations}
\eeq

\end{document}